\documentclass[journal]{IEEEtran}
\usepackage{amsthm}
\usepackage{amsmath}
\usepackage{amssymb}
\usepackage{graphicx}
\usepackage{url}
\usepackage{dsfont}
\usepackage{float}
\usepackage{bm}
\usepackage{ifthen}
\usepackage[usenames,dvipsnames]{color}
\usepackage{mathrsfs}
\usepackage[colorlinks=true,citecolor=blue,urlcolor=black]{hyperref}
\usepackage{float}
\usepackage{afterpage}
\usepackage{subfigure}
\usepackage{pifont}
\usepackage{color}

\newcommand{\be}{\begin{equation}}
\newcommand{\ee}{\end{equation}}


\newcommand{\sket}[1]{{\ensuremath{\lvert#1\rangle}}}
\newcommand{\lket}[1]{{\ensuremath{\left\lvert#1\right\rangle}}}
\newcommand{\ket}[1]{\if@display\lket{#1}\else\sket{#1}\fi}
\newcommand{\sbra}[1]{{\ensuremath{\langle#1\rvert}}}
\newcommand{\lbra}[1]{{\ensuremath{\left\langle#1\right\rvert}}}
\newcommand{\bra}[1]{\if@display\lbra{#1}\else\sbra{#1}\fi}
\newcommand{\sbraket}[2]{{\ensuremath{\langle#1\rvert#2\rangle}}}
\newcommand{\lbraket}[2]{{\ensuremath{\left\langle#1\!\left\rvert\vphantom{#1}#2\right.\!\right\rangle}}}
\newcommand{\braket}[2]{\if@display\lbraket{#1}{#2}\else\sbraket{#1}{#2}\fi}

\newcommand{\sketbra}[2]{{\ensuremath{\lvert #1\rangle\!\langle #2\rvert}}}
\newcommand{\lketbra}[2]{{\ensuremath{\left\lvert #1\right\rangle\!\!\left\langle #2\right\rvert}}}
\newcommand{\ketbra}[2]{\if@display\lketbra{#1}{#2}\else\sketbra{#1}{#2}\fi}

\theoremstyle{plain}

\theoremstyle{definition}

\begin{document}

\title{Measurement-device-independent quantum cryptography}

\author{Feihu Xu, Marcos Curty, Bing Qi, and Hoi-Kwong Lo
\thanks{F. Xu is with the Center for Quantum Information and Quantum Control, the
Department of Physics, and the Department of Electrical and Computer Engineering,
University of Toronto, Toronto, ON M5S3G4, Canada and also with
the Research Laboratory of Electronics, Massachusetts Institute of Technology,
Cambridge, MA 02139 USA.}
\thanks{Marcos Curty is with the Escuela de Ingenier\'{\i}a de Telecomunicaci\'{o}n, Department of Signal Theory and Communications, University of Vigo, Vigo, Pontevedra, E-36310, Spain.}
\thanks{Bing Qi is with the Quantum Information Science Group, Computational Sciences and Engineering Division, Oak Ridge National Laboratory, Oak Ridge, TN 37831-6418, USA.}
\thanks{H.-K. Lo is with the Center for Quantum Information and Quantum Control,
the Department of Physics, and the Department of Electrical and Computer
Engineering, University of Toronto, Toronto, ON M5S3G4, Canada.}
\thanks{Manuscript received ...; revised ....}}

\markboth{IEEE JOURNAL OF SELECTED TOPICS IN QUANTUM ELECTRONICS, VOL..., NO..., August 2014}%
{Shell \MakeLowercase{\textit{et al.}}: Bare Demo of IEEEtran.cls for Journals}

\IEEEspecialpapernotice{(Invited Paper)}

\maketitle

\begin{abstract}
In theory, quantum key distribution (QKD) provides information-theoretic security based on the laws of physics. Owing to the imperfections of real-life implementations, however, there is a
big gap between the theory and practice of QKD, which has been recently exploited by several
quantum hacking activities. To fill this gap,
a novel approach, called measurement-device-independent QKD (mdiQKD), has been proposed. It can
remove all side-channels from the measurement unit, arguably the most vulnerable part in QKD systems, thus offering
a clear avenue towards secure QKD realisations. Here, we review the latest
developments in the framework of mdiQKD, together with its
assumptions, strengths and weaknesses.
\end{abstract}

\IEEEpeerreviewmaketitle

\section{Introduction}
Secure communication is essential in today's digital society, with billions of users accessing the Internet via different terminals and mobile devices.
The goal is to transmit a secret message from a sender (Alice) to a receiver (Bob)
such that an eavesdropper (Eve) cannot access the message.
It is well known that this problem can be solved using the one-time-pad protocol~\cite{Vernam:OTP:1926}. This protocol
requires, however, that Alice and Bob share a secret key. With this key,
Alice can encrypt the message (so-called plaintext) into a ciphertext
that is unintelligible to Eve. On receiving the ciphertext,
Bob can recover the plaintext by using
his key. Importantly, the security of the one-time-pad protocol relies only on the secrecy of the shared key. This renders the
problem of secure communication essentially equivalent to that of distributing a key securely. This is the task of quantum key distribution (QKD)~\cite{bennett1984quantum}.

Unlike the widely-used public-key cryptography~\cite{RSA:1978}, which bases its security
on unproven computational assumptions, QKD can provide information-theoretically secure key distribution
based solely on the laws of physics.
Indeed, if a quantum computer is ever built, many classical public-key schemes will become insecure~\cite{Shor:1997}.
In sharp contrast, QKD will always remain secure despite the computational and technological power of Eve. That is,
when combined with the one-time-pad protocol, QKD can be used to achieve perfectly secure communication.

\subsection{Quantum key distribution (QKD)}
The best-known QKD protocol is the so-called BB84 scheme introduced by Bennett and Brassard in
$1984$~\cite{bennett1984quantum}. Its security is based on the quantum no-cloning
theorem, which states that it is impossible (for Eve) to make perfect copies of an unknown quantum state. Therefore,
the higher the amount of information that Eve learns
about a quantum signal, the higher the amount of
disturbance that she causes on it. By sacrificing a randomly chosen portion of their data, Alice and
Bob can estimate this disturbance (by calculating, for instance, the quantum bit error rate (QBER)) and thus bound Eve's information
about the distributed key. This compromised
information can then be removed from the final key by using privacy amplification methods.
The unconditional security of QKD has been rigorously proven in several papers~\cite{Mayers:2001, Lo:1999,Shor:2000,GLLP:2004,biham05}.

\subsection{Quantum hacking}\label{hacking}
While in theory QKD is unconditionally secure, in practice, however, there is an important gap between the assumptions made
in the security proofs of QKD and the actual implementations. This is so because real devices suffer from inevitable imperfections that can cause them to operate quite differently from
the mathematical models used to prove security. As a result, Eve could exploit such imperfections
to learn the distributed key
without being detected.

\begin{figure}[!t]
\centering
\resizebox{8.7cm}{!}{\includegraphics{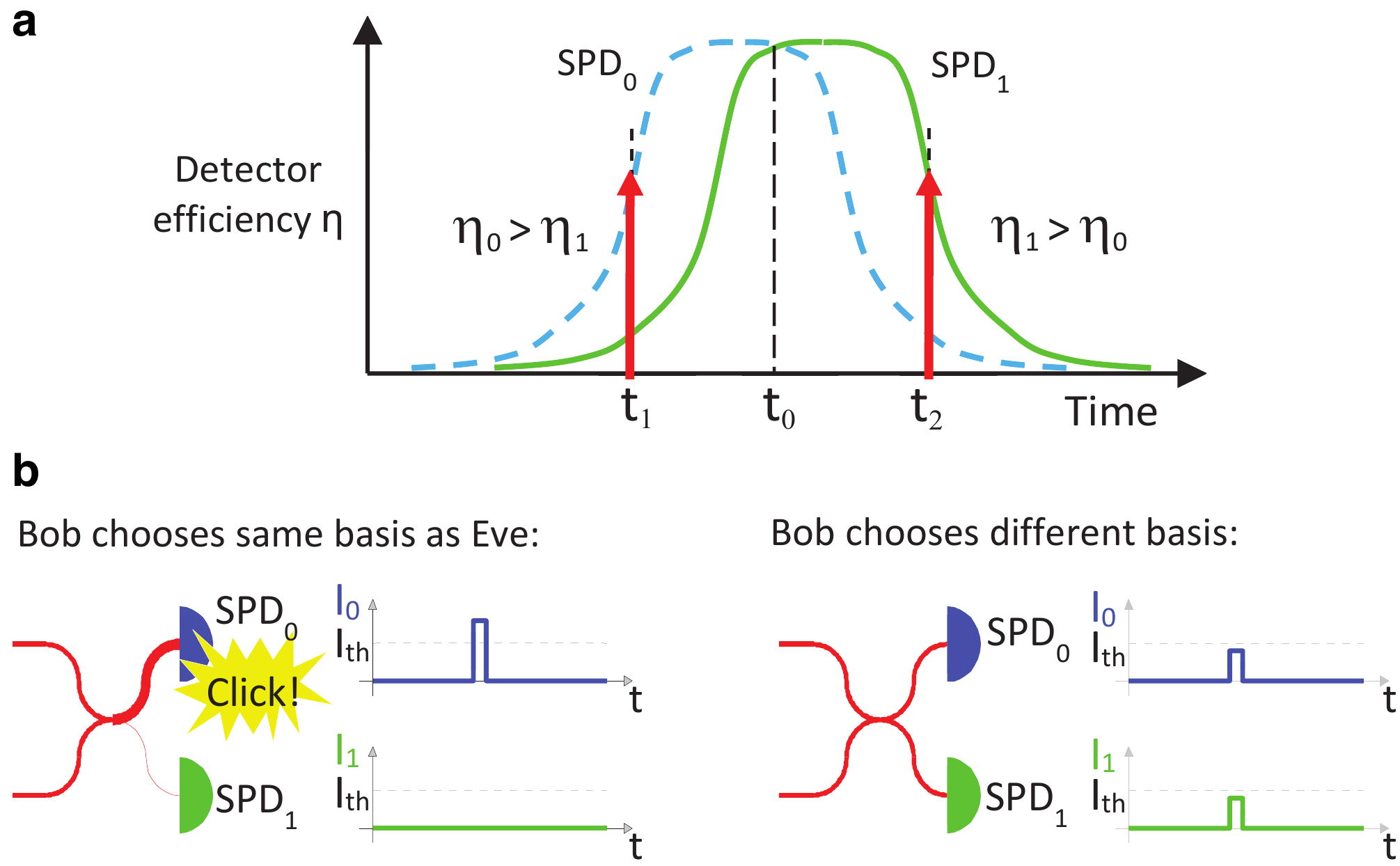}} \caption{(Color online) (a)
Schematic illustration of the typical detection efficiency mismatch between two single-photon detectors (SPD$_{0}$ and SPD$_{1}$ in the figure).
This could be used by Eve to perform a time-shift attack~\cite{Yi:timeshift:2008}. In this attack,
Eve shifts the arrival time of each signal sent by Alice to either $t_1$ or $t_2$ such that one detector has a much higher detection efficiency than the other.
In so doing, she can obtain information about the final key without being detected.
(b) Schematic illustration of the detector blinding attack~\cite{Lars:nature:2010}. First, Eve sends Bob bright light (not shown in the figure)
to blind his detectors and make them enter into linear mode operation. After that, she sends Bob a
tailored light pulse that produces a ``click'' in one of his detector only when he uses the same measurement basis employed by Eve to prepare her signal. Otherwise, no detector ``clicks''.
As a result, Eve can determine which detector generates a ``click'' at each given time and thus learn the whole key without introducing any
noticeable disturbance. (Figure 1b adapted with permission from~\cite{Lars:nature:2010}).
} \label{Fig:timeshift_blinding}
\end{figure}

The first successful quantum hacking attack against a commercial QKD system was the time-shift attack~\cite{Yi:timeshift:2008}.
It is based on an earlier theoretical
proposal introduced in~\cite{Qi:timeshift:2007} (see Fig.~\ref{Fig:timeshift_blinding}a).
Standard single-photon detectors (SPDs) such as, for example, InGaAs avalanche photodiodes, are often operated in a gated mode~\cite{Hadfield:2009}.
This means that their detection efficiency is time-dependent~\cite{Makarov:fakestate:2006}. Importantly,
since every QKD system contains at least two detectors
to measure two different bit values, it is usually quite difficult to guarantee that both detectors have precisely the same detection efficiency all the time.
In this scenario, Eve can simply shift the arrival time of each signal such that one detector has a
much higher detection efficiency than the other~\cite{Qi:timeshift:2007}. As a result, she could obtain partial information about the final key without introducing almost any error.

Recently, a more powerful attack---the so-called detector blinding attack---was introduced in~\cite{Lars:nature:2010}. It allows Eve to learn the whole key without being detected. The procedure is as follows (see Fig.~\ref{Fig:timeshift_blinding}b).
Eve sends bright light to Bob's detectors to force them enter into the so-called linear mode operation~\cite{Lars:nature:2010}. In so doing, the SPDs are no
longer sensitive to single-photon pulses but they behave like classical intensity photo-detectors. As a consequence, Eve can now fully control which detector ``clicks''
just by sending Bob a tailored light pulse. This attack has been successfully implemented against both commercial~\cite{Lars:nature:2010,yuan2010avoiding,Lars:avoiding:2010}
and research~\cite{Gerhardt:2010} QKD setups.

Quantum hacking has attracted a lot of research interest in recent years, and several attacks have been demonstrated experimentally~\cite{timeinfo,Xu:phaseremapping:2010,weier2011quantum,jain2011device,li2011attacking,tang2013source,jouguet2013preventing,jiang2013intrinsic,PhysRevLett.112.070503}.
Most of them exploit loopholes in Bob's measurement
system~\cite{timeinfo,weier2011quantum,jain2011device,li2011attacking,jouguet2013preventing,jiang2013intrinsic,PhysRevLett.112.070503},
which can be considered as the Achilles heel of QKD implementations. This is so because Eve can send Bob any signal she wishes,
which makes the protection of Bob's device quite difficult. In contrast, Alice can in principle protect her source by using, for instance, optical isolators.
It is therefore reasonable to expect that she can determine
the quantum states that she prepares and then include this information in the security analysis~\cite{kiyo_loss,xu2014experimental}. Note that a recent experiment has shown that the state preparation flaws do not significantly affect the performance of decoy-state BB84~\cite{xu2014experimental}.

\subsection{Countermeasures against quantum hacking}
Currently there are at least four main possible approaches to avoid the problem of quantum hacking and recover the security of QKD implementations.

The first solution consists in obtaining ``precise mathematical models'' for all the devices and then include this information into the
security proof~\cite{Fred:Proof:2009,PhysRevA.82.032337}. However, as QKD components
are complex apparatuses, this approach is unfortunately very challenging to realise in practice.

The second one is what we call ``patches''. Fortunately, once a particular attack is known, it is typically relatively easy to find
an appropriate countermeasure against it~\cite{ferreira2012real,yuan2011resilience}.
The main drawbacks of this solution are, however, unanticipated attacks. This is so because this approach only protects
against known hacking strategies. Therefore, its security resembles that of classical cryptography.

The third solution is called device-independent QKD (diQKD)~\cite{mayers_diQKD,acin2007device,diQKD3}.
Here, Alice and Bob do not need to know how their devices operate, but they can treat them as two ``black boxes''.
It requires, however, that certain assumptions are satisfied (see Table~\ref{Tab:assumptions}).
For instance, Alice and Bob need to guarantee that there is no leakage of unwanted information from their measurement apparatuses. In this scenario,
it is possible to prove the security of diQKD based solely on the
violation of a Bell inequality, which certifies the existence of quantum correlations.
Remarkably, this approach can remove all side-channels from QKD implementations. Its main
drawback, however, is that it needs a loophole-free Bell test which so far has never been performed. Also, the
expected secret key rate at practical distances is unfortunately very limited with current technology
(of the order of $10^{-10}$ bits per signal)~~\cite{qbitamp1,qbitamp2}. Still, diQKD could be a viable solution for short distance transmission in the
future with improved technology.
\begin{table}
\caption{Assumptions in diQKD versus mdiQKD}~\label{Tab:assumptions}
\begin{tabular}{l|c|c}
  \hline \hline
   \ & diQKD & mdiQKD  \\
     \hline
  True random number generators & Yes & Yes \\
   \hline
  Trusted classical post-processing & Yes & Yes \\
  \hline
  Authenticated classical channel & Yes & Yes \\
 \hline
  No unwanted information leakage
  from the & \ & \  \\
  measurement unit & Yes & {\it No}   \\
  \hline
  Characterised source & {\it No} & Yes  \\
  \hline
   \hline
\end{tabular}
\end{table}

In summary, the first solution is very difficult to realise in practice, the second one is ad hoc and
cannot provide information-theoretic security, and diQKD is unfeasible with present-day
technology. The rest of the paper is dedicated to analyse the fourth solution, which is
called measurement-device-independent QKD (mdiQKD)~\cite{Lo:MDIQKD}. As will be discussed below,
this approach can remove all side-channels from the measurement unit, which is (as discussed previously)
the weakest part of a QKD realisation. Most importantly, mdiQKD is fully
practical with current technology and it allows QKD with a high key rate and at a long distance. Therefore, mdiQKD offers a clear avenue to bridge the gap between theory and practice in QKD.

\section{Measurement-device-independent QKD}\label{mdiQKD}
\begin{figure}
\begin{center}
 \includegraphics[scale=0.4,angle=0]{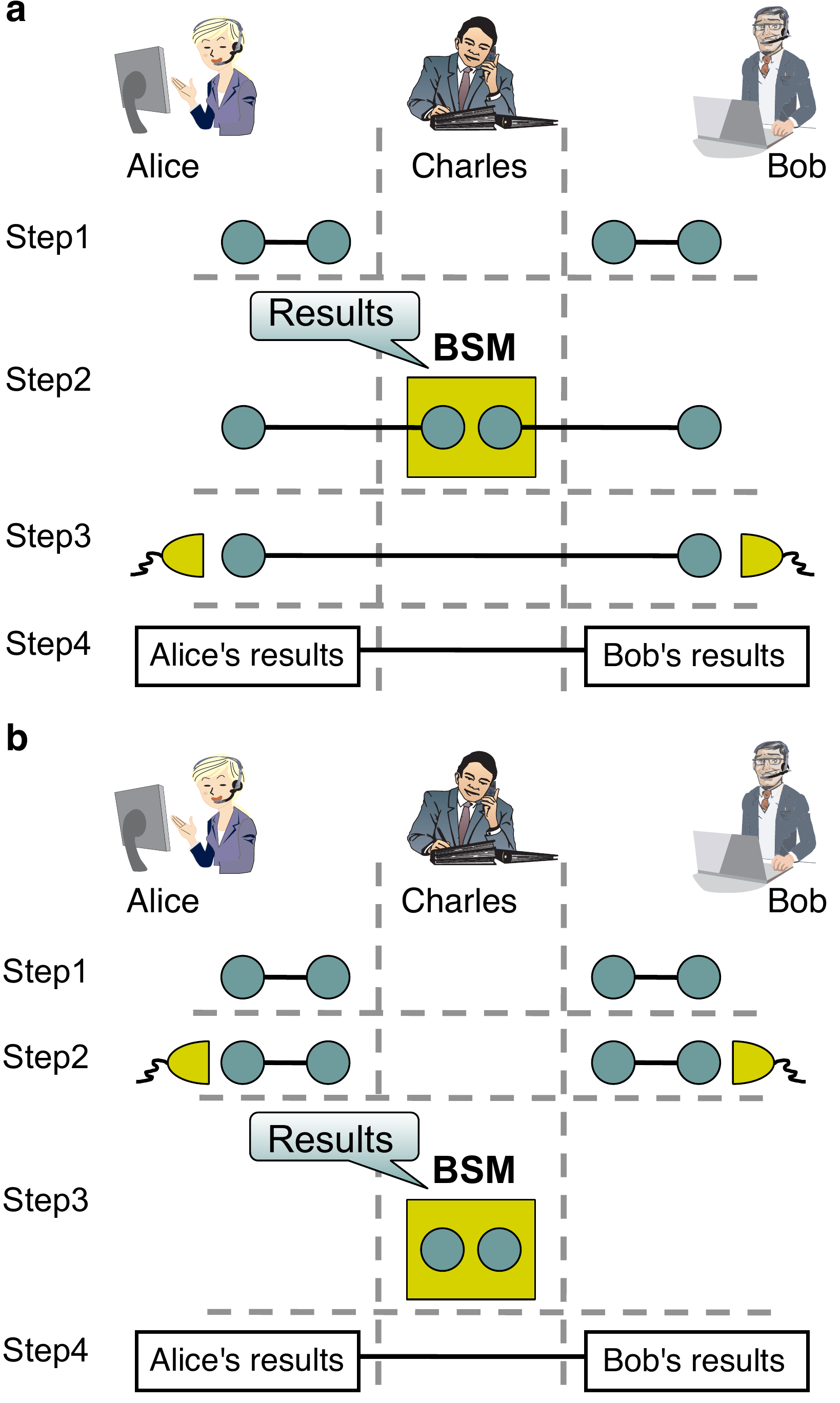}
 \end{center}
 \caption{(Color online) (a) An Einstein-Podolsky-Rosen (EPR) based QKD protocol. Step 1: Each of Alice and Bob prepares an EPR pair and sends half of it
 to an untrusted third party, Charles. Step 2: On receiving the signals, Charles is
supposed to realise an entanglement swapping operation via a Bell state measurement (BSM) and then
 broadcast his measurement results. Step 3: Alice and Bob measure their particles using the
 $\rm X$ or $\rm Z$ bases, which they select at random. Step 4: Alice and Bob test the honesty of Charles
 by comparing a randomly chosen portion of their data.
 (b) An equivalent time-reversed EPR based QKD protocol~\cite{biham1996quantum}. Since Charles' operations commute with those of Alice and Bob, one can reverse the order of the measurements (steps 2 and 3 in subfigure (a)). That is, Alice and Bob can safely measure their signals before Charles actually implements the BSM.
\label{Fig:EPRmdiQKD}}
\end{figure}

To understand the working principle of mdiQKD, let us first introduce an Einstein-Podolsky-Rosen (EPR) based QKD protocol (see Fig.~\ref{Fig:EPRmdiQKD}a).
Each of Alice and Bob prepares an EPR pair and sends half of it to an \emph{untrusted} third party, Charles. Charles is supposed to perform an entanglement swapping operation~\cite{zukowski1993event,pan1998experimental} on the incoming signals via a Bell state measurement (BSM), and then broadcast his measurement results. Once this step is completed, Alice and Bob measure their halves of the EPR pairs by using two conjugate bases (the rectilinear basis $\rm Z$, or the diagonal basis $\rm X$)
that they select at random. Importantly, by doing so they can determine whether or not Charles is honest. For this, they can compare a randomly chosen subset of
their data to test if it satisfies the expected correlations associated with the Bell state declared by Charles.

Interestingly, this protocol can also be implemented in a ``time-reversal'' fashion (see Fig.~\ref{Fig:EPRmdiQKD}b).
This is so because Charles' operations commute with those of Alice and Bob.
Therefore, one can reverse the order of the measurements. That is, it is not necessary that
Alice and Bob wait for Charles's results in order to measure their halves of the EPR pairs, but they
can measure them beforehand. Note that Charles'
BSM is only used to check the parity of Alice's and Bob's bits and, therefore,
it does not reveal any information about the individual bit values.
This rephrases the original EPR based QKD protocol into an equivalent prepare-and-measure scheme where
Alice and Bob directly send Charles BB84 states and Charles performs the measurements. Most importantly,
like in the original EPR based QKD protocol, Alice and Bob can test the honesty of Charles
by just comparing a random portion of their signals.

This time-reversal scenario has been studied in\footnote{Note that this approach also resembles
that proposed in~\cite{Lo:1999}, where Bob teleports any incoming signal from outside to himself by setting up a
teleportation gateway inside his own secured laboratory. In so doing, it is possible to remove all side-channels from his
measurement unit. However, note that the authors of~\cite{Lo:1999} considered that the BSM that is required for teleportation
could be trusted.}~\cite{biham1996quantum,inamori2002security} (see also~\cite{braunstein2012side} which
discussed the BSM conducted by Charles in QKD). Unfortunately, however, these important works offered very limited performance and, therefore,
they had been largely forgotten by the QKD community. For instance,
the scheme in~\cite{biham1996quantum} requires
perfect single-photon sources and long-term quantum memories, which renders it unpractical with current technology.
Inamori's scheme~\cite{inamori2002security} uses practical weak coherent pulses (WCPs) but it does not include decoy states, as
it was proposed long before the advent of the decoy-state method~\cite{Hwang:2003,Lo:2005, Wang:2005}.
The main merits of the mdiQKD proposal introduced in~\cite{Lo:MDIQKD} are twofold: first, it realised the importance of the
results in~\cite{biham1996quantum,inamori2002security}
to remove all detector side-channels
from QKD implementations and, second, it significantly improves the system performance with practical signals
by including decoy states.

\begin{figure}
\begin{center}
 \includegraphics[scale=0.8,angle=0]{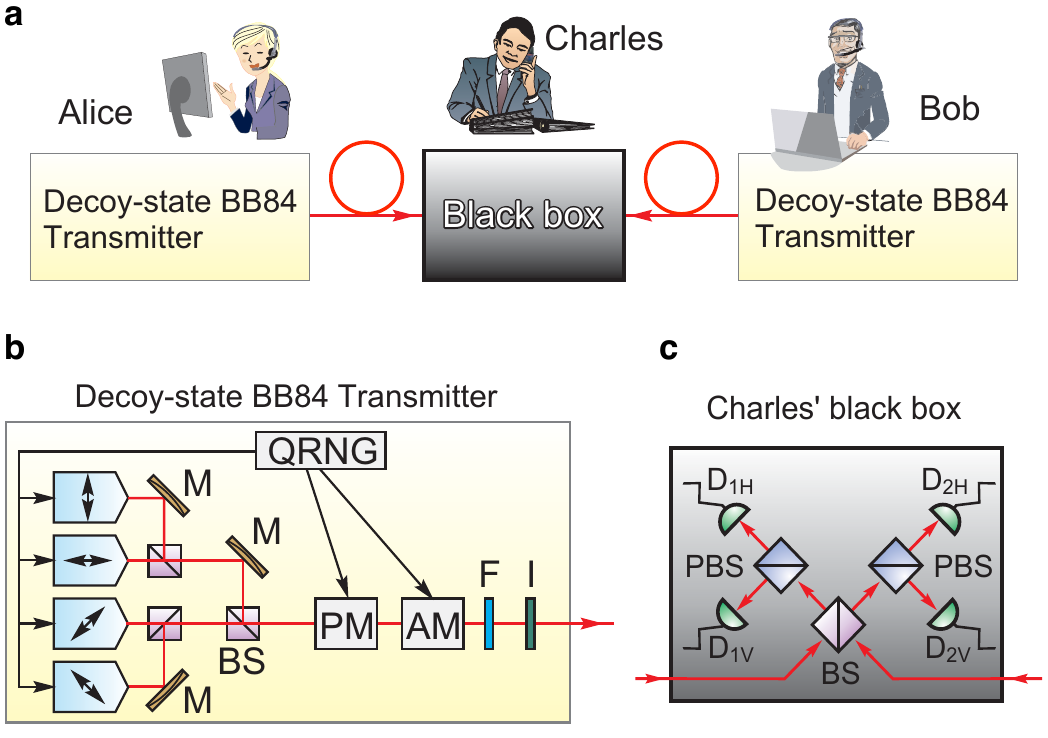}
 \end{center}
 \caption{(Color online) (a) Schematic diagram of a possible measurement-device-independent QKD (mdiQKD)
 implementation~\cite{Lo:MDIQKD}.
 Alice and Bob prepare BB84 polarisation states and
send them to an untrusted relay Charles/Eve, which can be treated as a ``black box''.
The relay is supposed to perform a Bell state measurement (BSM) that projects Alice's and Bob's signals into a Bell state.
In contrast to diQKD, note that now
there is no need to protect Charles' black box from any unwanted information leakage.
(b) Example of a decoy-state BB84 transmitter. The WCPs are generated using four emitting laser diodes. These
signals are then phase-randomised with a phase modulator (PM). Decoy states are prepared
using an amplitude modulator (AM). In the figure: (M) mirror, (BS) beam-splitter, (QRNG) quantum random number generator, (F) optical filter and (I) optical isolator. (c)
Example of a BSM implementation with linear optics.
Charles interferes the incoming pulses at a 50:50 BS, which has on each end a polarising beam-splitter (PBS) that projects the photons into
either horizontal ($H$) or vertical ($V$) polarisation states. A ``click'' in the single-photon detectors $D_{\rm 1H}$ and $D_{\rm 2V}$, or in $D_{\rm 1V}$ and $D_{\rm 2H}$,
indicates a projection into the singlet state $\ket{\psi^{-}}=(\ket{HV}-\ket{VH})/\sqrt{2}$, while a ``click'' in $D_{\rm 1H}$ and $D_{\rm 1V}$, or in $D_{\rm 2H}$ and $D_{\rm 2V}$, implies a projection into the triplet state $\ket{\psi^{+}}=(\ket{HV}+\ket{VH})/\sqrt{2}$. Other detection patterns are considered unsuccessful.
\label{Fig:mdiQKD}}
\end{figure}

An example of a possible mdiQKD implementation is illustrated in Fig.~\ref{Fig:mdiQKD}a~\cite{Lo:MDIQKD}. The protocol can be
summarised in the following three steps (see the caption of Fig.~\ref{Fig:mdiQKD} for further details):

{\it Step 1:} Alice and Bob prepare phase-randomised WCPs (together with decoy signals) in the BB84 states and send them to an untrusted relay Charles.

{\it Step 2:} If Charles is honest, he performs a BSM that projects Alice's and Bob's signals into a Bell state. In any case, he announces whether or not his measurements
are successful, including the Bell states obtained.

{\it Step 3:} Post-processing: Alice and Bob keep the data corresponding to Charles' successful measurement results and discard the rest. Also, they post-select the events where they employ the same basis and, based on the outcomes announced by Charles, say Alice flips part of her bits to correctly correlate them with those of Bob. Finally, they use the
decoy-state method to estimate the gain and QBER of the single-photon contributions\footnote{The idea of the decoy state protocol is that each of Alice and Bob uses an intensity modulator to modulate the intensity of each photon pulse. For instance, in a two decoy state protocol, the intensity of each pulse is chosen randomly from a set of prescribed intensities (with average photon numbers $\mu$, $\nu$ and $\omega$), where $\mu$ (e.g. about 0.5) is the signal state, $\nu$ (about 0.1) is the first decoy state and $\omega$ (about 0.005 or even 0) is the second decoy state. The key insight of decoy state is that, given a pulse of \emph{n} photons sent by Alice/Bob, Eve cannot differentiate it from the case where it has originated from a signal state or a decoy state. In other words, the gain and QBER depend only on the actual photon number \emph{n}, but are independent of whether it has originated from a signal or a decoy. After the transmission and Charles' broadcast of his measurement outcomes, Alice and Bob announce which pulses are signals and which pulses are decoys (i.e., the intensity, $\mu$, $\nu$ or $\omega$ used). Therefore, they can subsequently compute the gain and the QBER for all the combinations of signals and decoys. Based on such data, using the method described in Section IV-A, Alice and Bob can then estimate a lower bound to the single-photon contributions (together with their error rate) to Charles' successful BSM outcomes. From there, they can obtain a lower bound on the key generation rate.}.

Note that opposed to diQKD, now there is no need to protect Charles' measurement unit from unwanted leakage of information to the outside (see Table~\ref{Tab:assumptions}).
Indeed, this device can be fully controlled or even manufactured by Eve. This is significant because it means that there is no need to certify the detectors in a QKD standardisation process\footnote{The idea of using an untrusted relay (e.g. a BSM by Charles) in QKD has been discussed in not only~\cite{biham1996quantum,inamori2002security}, but also in diQKD proposals such as~\cite{braunstein2012side,qbitamp1,qbitamp2}. However, all these approaches have limited performance in terms of key rate and distance with current technologies. Compared to these solutions, the key advantage of mdiQKD is its strong performance in key rate and distance with only present-day technologies.}.
A small drawback of mdiQKD is that Alice and Bob need to know which states they send to Charles\footnote{The idea of using a trusted source (in combination with an untrusted measurement unit) in QKD has previously been discussed in, for example, the prepare-and-measure version of one-sided diQKD~\cite{branciard2012one}.  Nonetheless, one-sided diQKD requires a high detection efficiency and its performance is also limited. Comparatively speaking, mdiQKD does not require a high detection efficiency and it can achieve a much higher key rate and distance with current technologies.}. However, as we have seen in Sec.~\ref{hacking},
it is reasonable to expect that they can indeed characterise their sources and protect their state preparation processes from Eve's influence. Also, as will be discussed below in Sec.~\ref{further},
recent results show that a full source characterisation is not absolutely necessary for mdiQKD to work~\cite{kiyo_loss,yin2013measurement,yin2014mismatched}.

In the asymptotic scenario where Alice and Bob send Charles an infinite number of signals, the secret key rate has the form~\cite{Lo:MDIQKD}
\begin{equation} \label{Eqn:Key:formula}
    R\geq Q_{11}^{\rm Z}[1-H_{2}(e^{\rm X}_{11})]-Q^{\rm Z}_{\mu\mu}f_{e}(E^{\rm Z}_{\mu\mu})H_{2}(E^{\rm Z}_{\mu\mu}),
\end{equation}
where $Q_{11}^{\rm Z}$ is the gain ({\it i.e.}, the probability that Charles declares a successful result) when Alice and Bob send him
one single-photon each in the $\rm Z$ basis; $e^{\rm X}_{11}$ is the phase error rate of these single-photon signals; $Q^{\rm Z}_{\mu\mu}$ and
$E^{\rm Z}_{\mu\mu}$ represent, respectively, the gain and the quantum bit error rate (QBER) in the $\rm Z$ basis when Alice and Bob
send Charles WCPs of intensity $\mu$; $f_{e}\geq 1$ is the error correction inefficiency function, and $H_2(x)$=$-x\log_2(x)-(1-x)\log_2(1-x)$
is the binary entropy function.

Eq.~(\ref{Eqn:Key:formula}) assumes that Alice and Bob use the $Z$ basis for key generation and the $X$ basis for testing only~\cite{lo2005efficient}.
The term $Q_{11}^{\rm Z}H_{2}(e^{\rm X}_{11})$ corresponds to the information removed from the final key in the privacy amplification step of the protocol,
while $Q^{\rm Z}_{\mu\mu}f_{e}(E^{\rm Z}_{\mu\mu})H_{2}(E^{\rm Z}_{\mu\mu})$ is the information revealed by Alice in the error correction step.
The quantities $Q^{\rm Z}_{\mu\mu}$ and $E^{Z}_{\mu\mu}$ are directly measured in the experiment, while $Q_{11}^{\rm Z}$ and $e^{\rm X}_{11}$ can be estimated using the decoy-state method~\cite{Lo:MDIQKD}.

\section{Experimental demonstrations}\label{exp}
Up to date, four successful independent mdiQKD experimental realisations have already been reported~\cite{Tittel:2012:MDI:exp,da2012proof,Liu:2012:MDI:exp,zhiyuan:experiment:2013}.
Table~\ref{Tab:exp} includes a brief summary of their main features. In the POP
(proof-of-principle) demonstrations~\cite{Tittel:2012:MDI:exp,da2012proof},
both Alice and Bob send the same quantum state repeatedly without random selection of the encoding states or bases.
Therefore, no secret key is actually distributed between Alice and Bob.
In~\cite{Liu:2012:MDI:exp,zhiyuan:experiment:2013}, however,
two real demonstrations with key exchange have been performed. Below, we review in more detail these results.

\begin{table}
\caption{Comparison between mdiQKD demonstrations}~\label{Tab:exp}
\begin{tabular}{c|c|c|c|c}
  \hline \hline
   & Ref.~\cite{Tittel:2012:MDI:exp} & Ref.~\cite{da2012proof} & Ref.~\cite{Liu:2012:MDI:exp} & Ref.~\cite{zhiyuan:experiment:2013}  \\
  \hline
  Encoding & Time-bin & Polarisation & Time-bin & Polarisation\\
    \hline
  Implementation & POP$^{*}$ & POP & Real$^{**}$ & Real \\
  \hline
  Condition & Field test  & Lab & Lab & Lab \\
    \hline
  Fiber length & 28.8 km & 17 km & 50 km & 10 km  \\
  \hline
  Asymp-key rate$^{\dag}$ & $\sim10^{-6}$  & $\sim10^{-6}$ &  N.A.  & $\sim10^{-6}$\\
  \hline
  Finite-key rate$^{\ddag}$ & N.A.$^{***}$  & N.A. & $\sim10^{-7}$ & $\sim10^{-8}$ \\
  \hline
  Repetition rate & 2 MHz & 1 MHz & 1 MHz & 0.5 MHz \\
  \hline \hline
\multicolumn{5}{l}{\textsuperscript{*}\footnotesize{POP represents ``proof-of-principle'' experiment, {\it i.e.}, no key exchange.}}\\
\multicolumn{5}{l}{\textsuperscript{**}\footnotesize{Real: there is true key exchange.}} \\
\multicolumn{5}{l}{\textsuperscript{***}\footnotesize{N.A. represents ``not available''}} \\
\multicolumn{5}{l}{\textsuperscript{\dag}\footnotesize{The key rate (bits/pulse) under the assumption of infinite long keys.}}\\
\multicolumn{5}{l}{\textsuperscript{\ddag}\footnotesize{The key rate (bits/pulse) after finite-key corrections.}}
\end{tabular}
\end{table}

The main experimental challenge of mdiQKD is to perform a high-fidelity BSM between photons from different light sources,
which is not required in conventional QKD schemes. Indeed,
to obtain high-visibility two-photon interference (and, therefore, to have a low QBER),
Alice's photons should be indistinguishable from those of Bob\footnote{In principle, one may suggest that first Charles sends strong light pulses (from the same laser) to both Alice and Bob,
who then encode their bit values, attenuate the pulses, and send them back to Charles. However, note that this design could compromise the security of the whole system,
as now Charles could try to interfere with Alice's and Bob's state preparation processes.}.
Furthermore, if one implements mdiQKD over telecom fibres, it is necessary to include feedback controls
to compensate the time-dependent polarisation rotations and propagation delays caused by the fibres.
Note that in standard QKD systems, this requirement can be relaxed by using
phase encoding~\cite{dixon2008gigahertz,rosenberg2009practical}, because the two optical pulses, which interfere with each other at the receiver's end, pass through the same optical fibre and thus experience the same polarisation rotation and phase change. Therefore, one can achieve high interference visibility without performing any polarisation control.
Nevertheless, this advantage of phase encoding (in comparison to other encoding schemes) cannot be directly translated to mdiQKD.
The reason is that now Alice's and Bob's signals are generated from two independent lasers and they propagate through two independent quantum channels. Consequently, in mdiQKD, polarisation management is required in all encoding schemes.

The feasibility of generating indistinguishable photons from two independent lasers was already investigated in the original
mdiQKD proposal~\cite{Lo:MDIQKD}, where the authors conducted a Hong-Ou-Mandel (HOM) experiment using two independent commercial off-the-shelf lasers. By carefully matching the central frequencies, the pulse shapes, the arrival times, and the polarisation states of the photons, an HOM interference visibility close to the theoretical value of 50\% was observed at different average photon numbers (from 0.2 to 4). The near perfect HOM visibility implied that the photons generated by the two lasers were almost identical. The same method was also adopted in more recent studies to verify the indistinguishability of photons from different sources~\cite{Tittel:2012:MDI:exp,da2012proof,Liu:2012:MDI:exp,zhiyuan:experiment:2013}.

So far, both time-bin~\cite{Tittel:2012:MDI:exp,Liu:2012:MDI:exp} and polarisation
encoding~\cite{da2012proof,zhiyuan:experiment:2013} mdiQKD have been demonstrated. In the first POP mdiQKD demonstration~\cite{Tittel:2012:MDI:exp}, an HOM interference experiment was conducted with photons generated by independent sources that travel through separate field-deployed fibres of lengths 6.2 km and 12.4 km, respectively.
By performing automatic polarisation stabilisation, manual adjustment of the photons arrival time, and manual adjustment of the lasers frequency,
a high interference visibility was obtained even under a real-world environment. The two light pulses constructing a time-bin signal were generated by modulating the output of a CW laser twice.
When compared to other recent experimental implementations of mdiQKD~\cite{da2012proof,Liu:2012:MDI:exp,zhiyuan:experiment:2013},
the scheme in~\cite{Tittel:2012:MDI:exp} required higher laser frequency stability. This is so because the phase difference between Alice's and Bob's lasers must be constant within a time window corresponding to two time-bin pulses rather than to only a single laser pulse.

\begin{figure*}[t]
\centerline{\includegraphics[width=13cm]{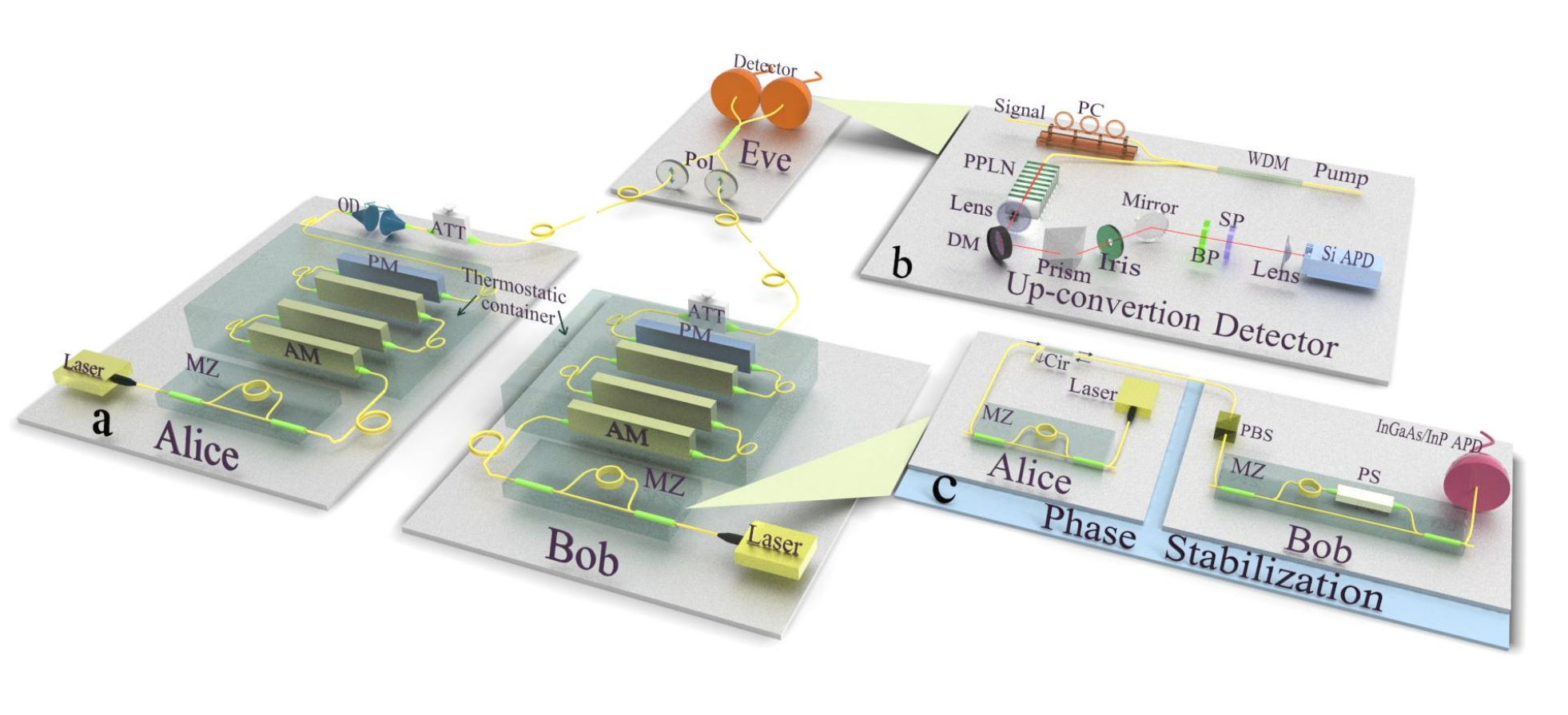}}
\caption{(Color online) Demonstration of mdiQKD using time-bin encoding realised in China~\cite{Liu:2012:MDI:exp}. (a) Alice (Bob) passes her (his) laser pulses through an unbalanced Mach-Zehnder (MZ) interferometer to generate two time-bin pulses. Three amplitude modulators (AMs) and a phase modulator (PM) are used to generate decoy states and encode the time-bin qubit. The pulses are then attenuated by an
attenuator (ATT) and pass through 25 km fibre spools of each arm to the measurement unit. The pulses finally interfere at a 50:50 fibre beam-splitter (BS) to perform a BSM. The two outputs are detected by up-conversion detectors. (Part b) Schematic diagram of the up-conversion single-photon detector. In the figure: (PC) polarisation controller, (DM) dichroic mirror, (BP) band pass filter, and (SP) short pass filter. (Part c) Schematic diagram of the phase stabilisation setup. (Cir) circulator, (PS) phase shifter, and (PBS) polarising beam-splitter. (Reprinted Figure with permission from~\cite{Liu:2012:MDI:exp}).}\label{Fig:MDItimebin}
\end{figure*}

A similar time-bin encoding mdiQKD scheme was investigated in~\cite{Liu:2012:MDI:exp} (see Fig.~\ref{Fig:MDItimebin}). The authors performed a real mdiQKD demonstration with random selection of encoding states and bases.  They made use of custom-made and specialised devices including high-efficiency up-conversion single-photon detectors. Here, the time-bin signals were generated by sending a laser pulse through an unbalanced fibre interferometer. Compared to~\cite{Tittel:2012:MDI:exp}, this approach relaxes the requirement on the frequency stability of two lasers but it needs phase stabilisation of the fibre interferometers.

\begin{figure*}[t]
\centerline{\includegraphics[width=13cm]{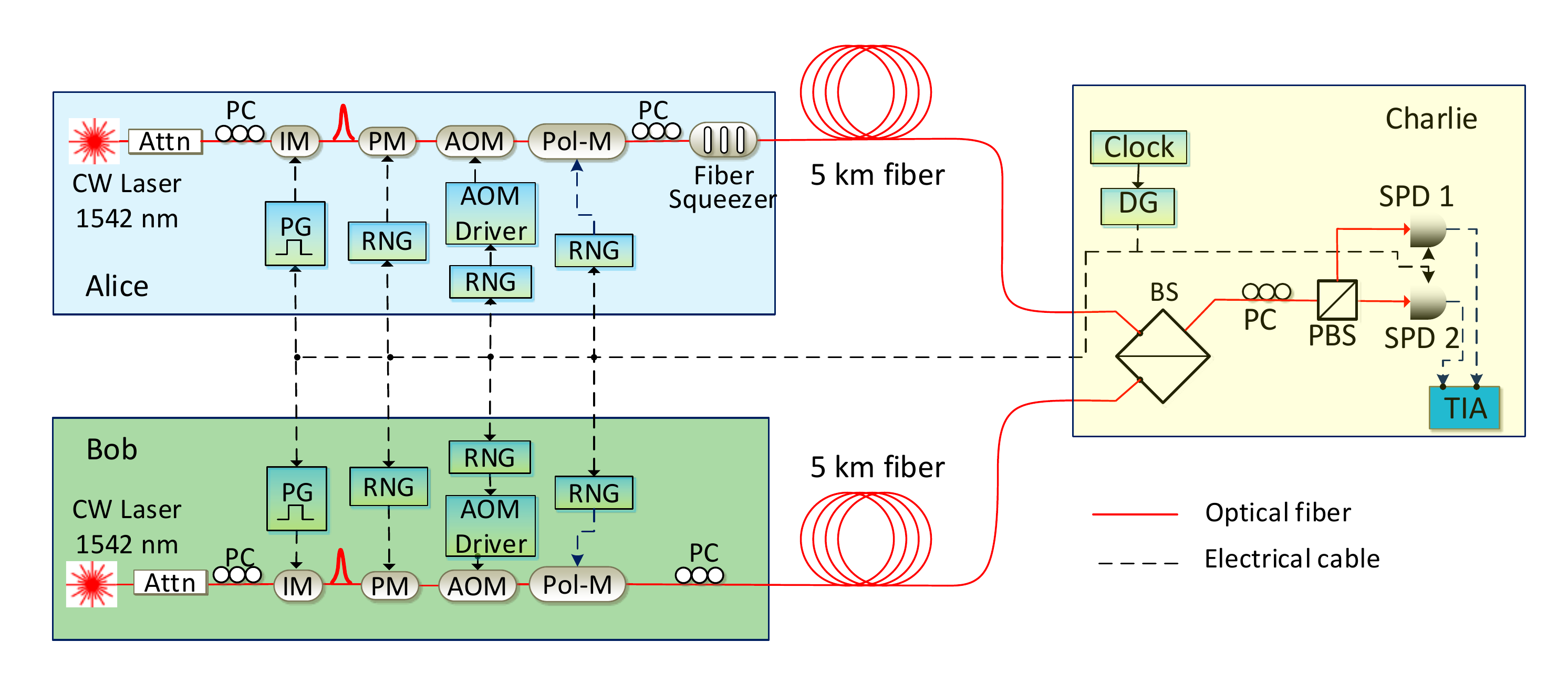}}
\caption{(Color online) Demonstration of mdiQKD using polarisation encoding realised in Toronto, Canada~\cite{zhiyuan:experiment:2013}.
Each of the two CW frequency-locked lasers is attenuated by an optical attenuator (Attn) and modulated by an intensity modulator (IM), which is driven by an electrical pulse generator (PG), to prepare weak coherent pulses (WCPs). The phases of these pulses are uniformly modulated by a phase modulator (PM) for active phase randomisation. Next, an acousto-optic modulator (AOM) randomly modulates
their intensities to implement the decoy-state protocol. Key bits are encoded into polarisation states of the WCPs using a polarisation modulator (Pol-M). Alice and Bob send their signals to Charles through a 5 km fibre spool. On receiving the transmission, Charles
performs a BSM on the incoming pulses using a beam-splitter (BS) and a polarising beam-splitter (PBS) together with two commercial SPDs. Synchronisation is done with the help of an electrical delay generator (DG). In the figure: (PC) polarisation controller, (RNG) random number generator and (TIA) time interval analyser.
(Figure reproduced from~\cite{zhiyuan:experiment:2013}).}\label{Fig:MDIpolarisation}
\end{figure*}

Polarisation encoding mdiQKD was also performed by two independent research groups. In particular,~\cite{da2012proof} demonstrated that
the polarisation rotation due to a long fibre could be compensated using conventional polarisation feedback control.~\cite{da2012proof} is a POP experiment and no key was actually exchanged. In~\cite{zhiyuan:experiment:2013}, the authors performed a real mdiQKD demonstration over 10 km of single-mode fibre using solely
commercial off-the-shelf devices. This implementation setup is shown in Fig.~\ref{Fig:MDIpolarisation}. The two lasers used in this experiment are two commercial frequency-stabilized lasers where the frequency of each laser is individually locked to a local gas cell that is integrated by the manufacturer. Thus, there is no optical or electronic link between the two lasers. Here, a finite-key security analysis~\cite{curty2014finite} was applied to optimise the experimental parameters and to evaluate the final secure key rate. This rate is slightly lower than that in~\cite{Liu:2012:MDI:exp} due to the limited detection efficiency and repetition rate of the system.

All these experiments, when taken together, complete the cycle needed to demonstrate the feasibility of using off-the-shelf optoelectronic devices to build a QKD system that is immune to all detector side-channel attacks.

\section{Theoretical aspects for mdiQKD}
As we have seen in Sec.~\ref{mdiQKD}, since practical and efficient single-photon sources are still unavailable, the original theoretical proposal of mdiQKD~\cite{Lo:MDIQKD},
together with the experimental realisations presented in the previous section,
 considers WCPs and decoy states instead. However, in order to apply the results of~\cite{Lo:MDIQKD} to real-life systems,
 there are still some loose ends that need to be addressed. In particular,
\cite{Lo:MDIQKD} assumes that Alice and Bob use an infinite number of decoy-state settings. Also, for simplicity, this study neglects statistical fluctuations due to
the finite data size. In this section, we review the latest developments in mdiQKD to overcome these limitations.

\subsection{Decoy-state protocol with a finite number of settings}

The analysis of this scenario is basically the same as that of conventional decoy-state QKD systems~\cite{Hwang:2003,Lo:2005, Wang:2005}. The sole difference is that now both Alice and Bob send decoy signals to a common receiver (instead of only Alice sending decoy states to Bob),
which makes the mathematics slightly more cumbersome. Fortunately, it has been shown that also in this situation it is enough if Alice and Bob
employ just two decoy settings each. That is, this configuration can already provide them with a quite tight estimation of the
relevant parameters that are needed to prove security ({\it i.e.}, $Q_{11}^{\rm Z}$ and $e^{\rm X}_{11}$ in
Eq.~(\ref{Eqn:Key:formula})).

For instance, the authors of~\cite{ma2012statistical} proposed a numerical method (based on linear programming) for the case where Alice and Bob employ
two (or three) decoy states each. Similarly, Refs.~\cite{wang2013three,sun2013practical} presented different analytical estimation
approaches, based on Gaussian elimination, under the assumption that Alice and Bob can prepare a vacuum state. Also,
following similar analytical lines, the authors of~\cite{Feihu:practical,curty2014finite} studied the situation where none of the two decoy signals are vacuum,
because a vacuum state is relatively hard to realise in practice due to the finite extinction ratio of a practical intensity modulator~\cite{rosenberg2009practical}.
More recently,~\cite{PhysRevA.89.052333} compared different decoy-state methods for mdiQKD and confirmed
that two decoy settings are enough to obtain a near optimal estimation of the relevant parameters. That is, the use of
three or more decoy intensities does not result in any significant improvement of the final secret key rate
in neither the asymptotic nor the finite data size regimes. All these results provide experimentalists a clear path
to implement mdiQKD with WCPs and decoy states.

\subsection{Finite-key security analysis}
The second question that needs to be solved is related with the fact that any QKD realisation only produces a finite amount of data.
Of course, a real-life QKD experiment is always completed in finite time, which means that the length of the output secret key is
obviously finite. Thus, the parameter estimation procedure in QKD needs to take the statistical fluctuations of the different parameters into account.
This problem has attracted a lot of research attention in recent years, and several security proofs in the finite-key regime for conventional QKD
systems have been obtained~\cite{renner2005security,scarani2008quantum}
(see also~\cite{QKD:Scarani:2009} for a review on this topic). Very recently, it was possible to obtain tight finite-key security bounds
for both the BB84 protocol with single-photons~\cite{tomamichel2012tight} and the decoy-state BB84 scheme~\cite{hayashi2014security,lim2014concise}. These security bounds are valid
against the most general attacks.

\begin{figure}[t]
\centerline{\includegraphics[width=0.51\textwidth]{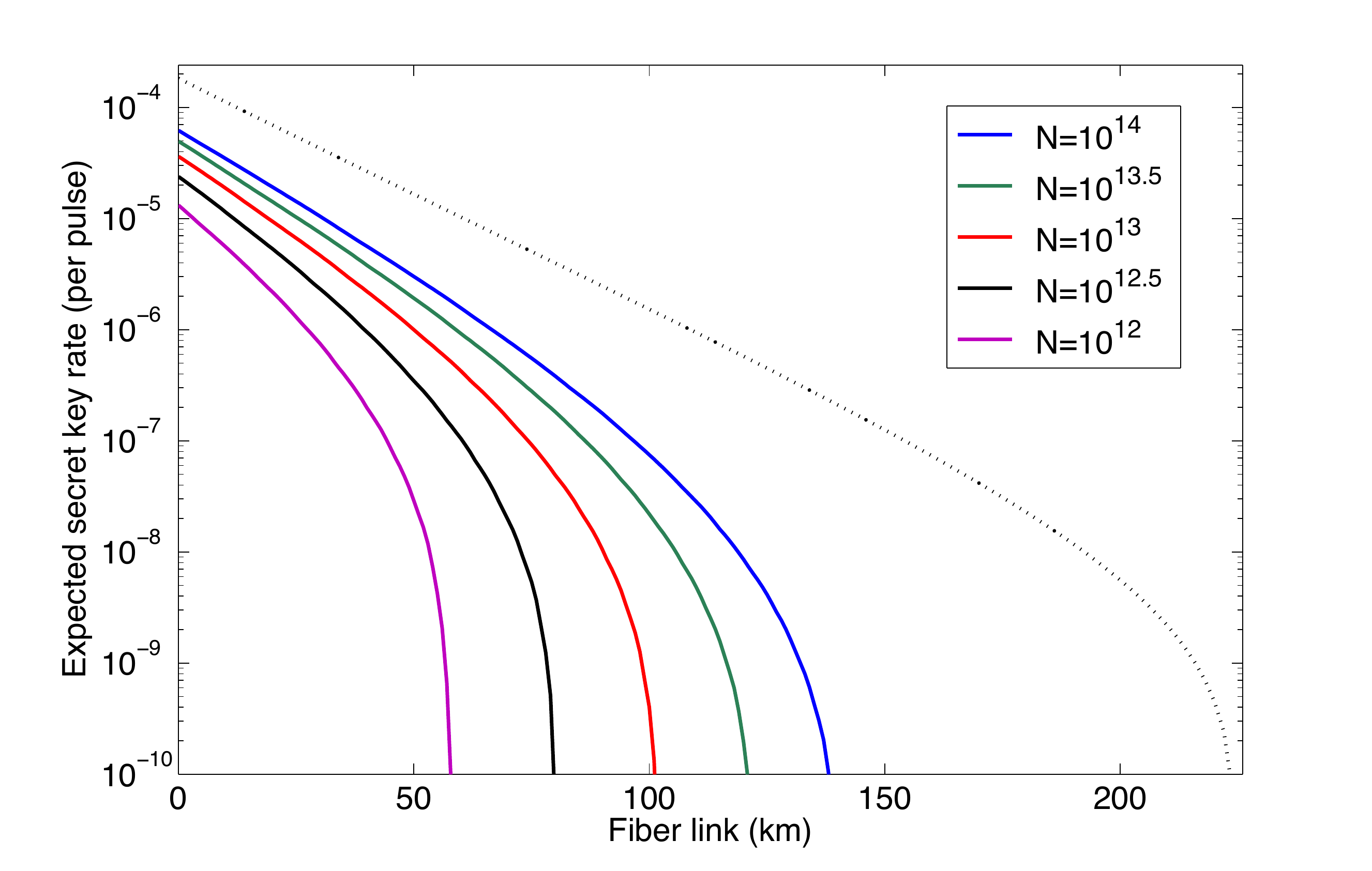}}
\vspace{-0.2cm}
\caption{(Color online) Secret key rate in logarithmic scale for mdiQKD as a function of the distance between Alice and Bob~\cite{curty2014finite}. The solid lines correspond to different values for the total number of signals $N$ sent by Alice and Bob. The security bound is $\epsilon=10^{-10}$. For simulation purposes, the following experimental
parameters are considered: the overall misalignment in the channel is $1.5\%$, the loss coefficient of the channel is $0.2$ dB/km, the detection efficiency of the relay is $14.5\%$, and the background count rate is $6.02\times{}10^{-6}$. These results demonstrate that even with a realistic finite size of data, say $N=10^{12}$ to $10^{14}$, it is possible to achieve secure mdiQKD at long distances. In comparison, the dotted line represents a lower bound on the secret key rate for the asymptotic case where Alice and Bob send Charles infinite signals and use an infinite number of decoy settings. (Figure reproduced from~\cite{curty2014finite}).
}\label{fig1:finitekey}
\end{figure}

Similar techniques can also be applied to mdiQKD. For example, the authors of~\cite{ma2012statistical} made a first step in this
direction and provided a finite-key security analysis that assumes a Gaussian distribution for
the statistical fluctuations. Also,~\cite{SongPhysRevA.86.022332} includes an analogous study that is valid
against particular types of attacks. More recently,~\cite{curty2014finite} presented a finite-key security proof that is valid against
general attacks and it does not assume any particular distribution for the statistical fluctuations. In addition, this result
satisfies the ``composable'' definition of the security of QKD~\cite{ben2005universal,renner2005universally}.
That is, the generated secret keys remain secure when they are
employed as a resource for other cryptographic systems ({\it e.g.}, the one-time-pad protocol). All these results
confirm the feasibility of long-distance implementations of mdiQKD (for instance, say 100 km of optical fibre with 0.2 dB/km loss)
with current technology and within a reasonable time-frame of signal transmission.
As an illustration, Fig.~\ref{fig1:finitekey} shows the simulation result presented in~\cite{curty2014finite}.

\subsection{Further theoretical developments}\label{further}
\subsubsection{Encoding schemes}
The original theoretical proposal of mdiQKD~\cite{Lo:MDIQKD} uses polarisation encoding. However, mdiQKD can be realised as well,
as we have seen in Section~\ref{exp}, using other encoding schemes like phase encoding~\cite{tamaki2012phase,ma2012alternative} or time-bin encoding~\cite{chan2014modeling}. Polarisation encoding is typically more suitable for free-space implementations due to the negligible birefringence of air, while phase encoding and time-bin encoding are usually more convenient for fibre-based realisations. In order to
select the optimal experimental parameters for these implementations, detailed theoretical system models have been developed in~\cite{Feihu:practical} (for
polarisation encoding), in~\cite{ma2012alternative} (for phase encoding), and in \cite{chan2014modeling} (for time-bin encoding). Also,
a parameter optimisation method has been presented in~\cite{PhysRevA.89.052333}.

Of course, the channels that connect Alice and Bob with Charles are typically of different length and have different transmittances. Indeed, such
asymmetric case is expected in most mdiQKD realisations~\cite{Tittel:2012:MDI:exp}; it has been studied in~\cite{Feihu:practical}.

\subsubsection{Extending the covered distance and achievable secret key rate}
\begin{figure}[t]
\centerline{\includegraphics[width=0.51\textwidth]{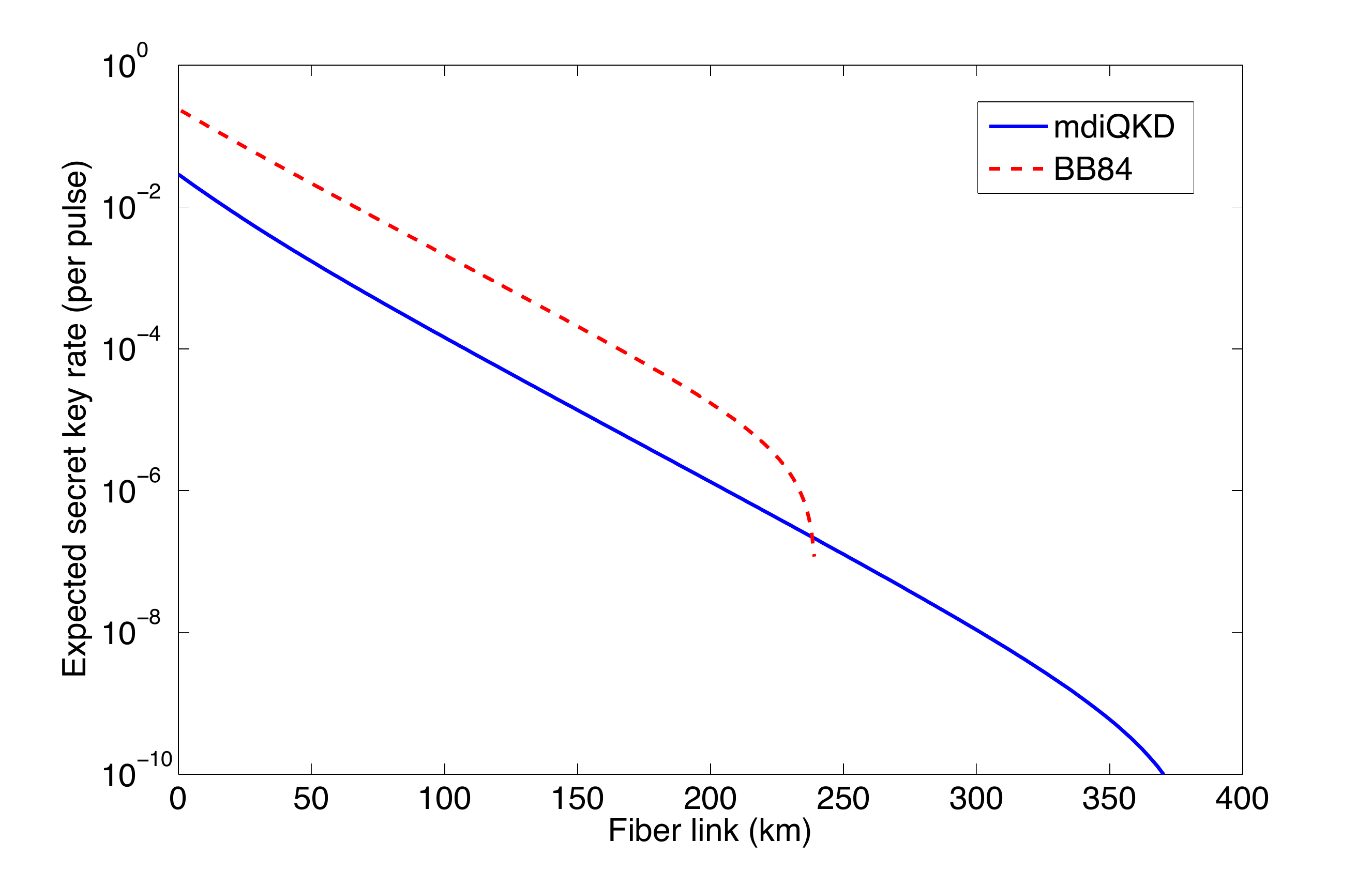}}
\vspace{-0.2cm}
\caption{(Color online) Asymptotic secret key rate in logarithmic scale for decoy-state BB84 and mdiQKD as a function of the distance between Alice and Bob. The detection efficiency of the SPDs is 93\%~\cite{marsili2013detecting} and other experimental parameters coincide with those of Fig.~\ref{fig1:finitekey}.}\label{fig:asympbetter}
\end{figure}

One possible solution to achieve these goals is, of course, to use
ultra-low loss fibres~\cite{stucki2009high} in combination with high detection efficiency SPDs~\cite{marsili2013detecting}. For instance, Fig.~\ref{fig:asympbetter} shows the asymptotic
secret key rate versus distance for both the decoy-state BB84 and mdiQKD, when the detection efficiency of the SPDs is 93\%~\cite{marsili2013detecting}. Other experimental parameters coincide with those of Fig.~\ref{fig1:finitekey}.

\begin{figure}[t]
\centerline{\includegraphics[width=8.8cm]{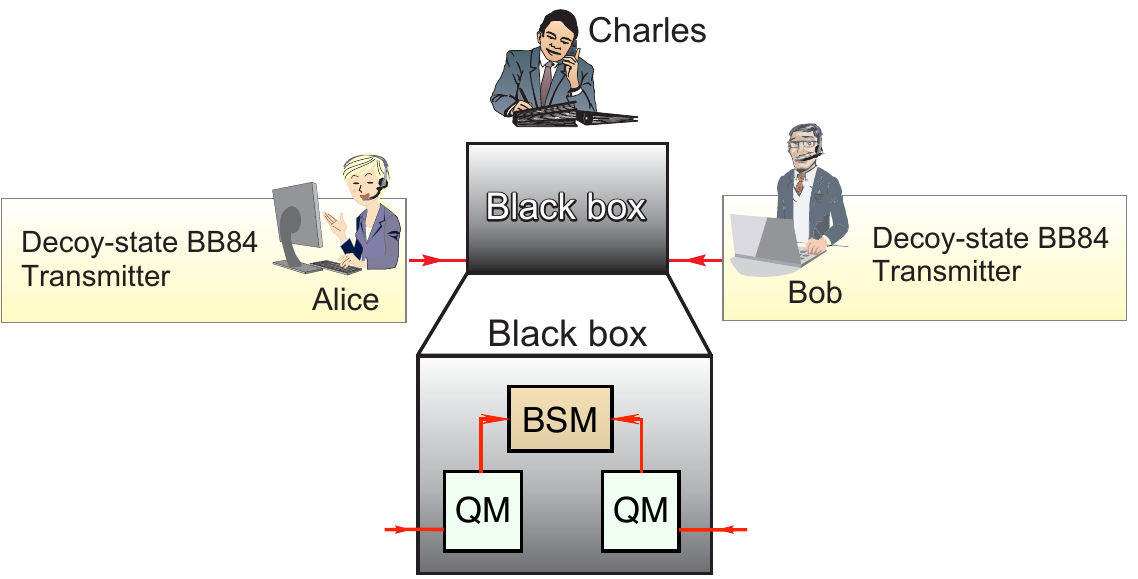}}
\caption{(Color online) Schematic diagram of mdiQKD with heralded quantum memories~\cite{abruzzo2014measurement,panayi2014memory}.
Charles first stores the incoming photons in two heralded quantum memories, one for Alice and one for Bob. Then, he
performs a Bell state measurement only between those photons that have been successfully stored in the memories. This
significantly increases the success probability of his measurement unit and, therefore, also the covered distance and achievable secret
key rate are increased.
In the figure: (QM) quantum memory, and (BSM) Bell state measurement. }\label{Fig:MDIQM}
\end{figure}

Alternatively, one could also include quantum memories in
Charles's measurement device. This last situation has been analysed in~\cite{abruzzo2014measurement,panayi2014memory} (see Fig.~\ref{Fig:MDIQM}).
The main idea is quite simple. Instead of performing a BSM between each pair of signals
received from Alice and Bob, Charles firstly stores the incoming photons in two heralded quantum memories, one for Alice's signals and one for Bob's signals. After that,
he performs a BSM only between those photons that have been successfully stored in the memories. By doing so, he can increase the
success probability of the measurement unit, which results in a significant increase of both the covered distance and the secret key rate.
Unfortunately, however, this type of approach is very challenging with current technology.

A more feasible solution nowadays is to place an entanglement source between two quantum relays. This scenario has been studied in~\cite{xu2013long}.
Its main drawback, however, is its limited key rate. This is so because of the low detection efficiency of today's SPDs~\cite{Hadfield:2009} together with the
fact that now one requires that two different BSMs are successful at the same time.

\subsubsection{State preparation flaws}
Recently, there have been also efforts to prove the security of BB84 and mdiQKD when Alice's and Bob's devices are flawed~\cite{kiyo_loss,xu2014experimental}, or when their apparatuses are not fully characterised~\cite{yin2013measurement,yin2014mismatched}. Remarkably,
the authors of~\cite{kiyo_loss,xu2014experimental} showed that state preparation flaws do not significantly affect the performance of BB84 and
mdiQKD. That is, it is not necessary that Alice's and Bob's state preparation process is very precise to obtain a good performance.
Indeed, a modified version of mdiQKD where Alice and Bob send Charles only three different states can deliver
the same key rate as the original scenario where they send him four BB84 states~\cite{kiyo_loss}.

\subsubsection{Alternative system implementations}
The idea of mdiQKD is compatible as well with other QKD protocols like continuous-variable schemes~\cite{li2014continuous,pirandola2013continuous,PhysRevA.89.042335}
or the Scarani-Acin-Ribordy-Gisin scheme~\cite{mizutani2014measurement}. Also, it can be implemented with different types of sources~\cite{PhysRevA.88.052332,li2014measurement}.

\section{Outlook}

On the experimental side, it would be necessary to improve the performance of the mdiQKD implementations realised so far. For instance,
current experimental demonstrations
consider short-distance transmission ({\it i.e.}, below 50 km) and their system clock rate is relatively low
(below 2 MHz). For practical applications, it would be desirable to achieve longer distances (say around 100-200 km)
and to use higher system clock rates (say 100MHz-1GHz)\footnote{After the completion of a preliminary version of this manuscript,
a new mdiQKD implementation over 200 km of optical fibre using a system clock of 75 MHz has been reported in~\cite{new_mdiQKDexp}. See also~\cite{6920009} for a field test.}. Using state-of-the-art SPDs~\cite{marsili2013detecting}, for example, could also help to substantially increase the final key generation rate.

\begin{figure}[t]
\centerline{\includegraphics[width=8.8cm]{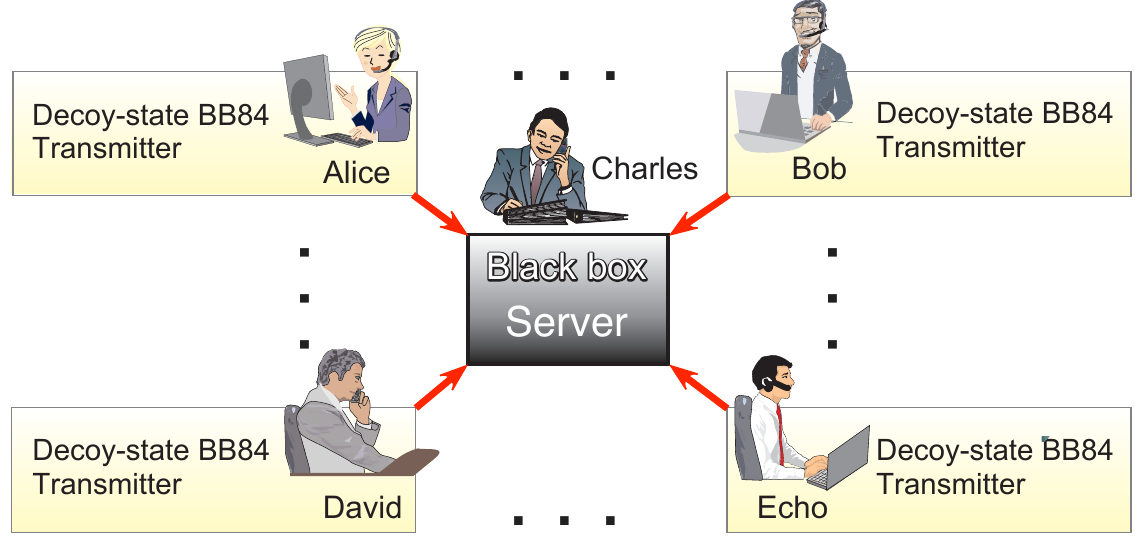}}
\vspace{-0.2cm}
\caption{(Color online) Example of a QKD network with an untrusted node based on mdiQKD. The users have low-cost and compact transmitters that send quantum signals
to the network server, which contains all the expensive and complex calibration and measurement devices. This scenario can be easily extended to the case
with several servers.}\label{Fig:MDInetwork}
\end{figure}

It would be interesting as well to prove the feasibility of mdiQKD for free-space communications.
Such implementation would constitute a first step towards future satellite-based mdiQKD networks, in which an untrusted satellite can be shared by many users.
Moreover, continuous-variable mdiQKD demonstrations using standard telecom devices are still missing. Furthermore, in the long term, mdiQKD could be used to build a fibre-based QKD network with untrusted nodes, in which the users possess low cost, compact devices to transmit quantum states, while all the expensive calibration and measurement apparatuses are located within the network servers\footnote{China and the US have announced that they are building long-distance quantum network based on trusted relays. China's network from Shanghai to Beijing will be 2000 km. A key weakness in such network is that one has to trust all the relays. MDI-QKD is advantageous as it will completely remove such trust.}. This scenario, illustrated in~Fig.~\ref{Fig:MDInetwork}, is advantageous over the recent demonstrations on QKD networks~\cite{frohlich2013quantum,hughes2013network}, as it completely removes the trust on the central relay node.

Much work needs to be done as well on the theoretical side. For instance, as already discussed,
a key assumption in mdiQKD is that Alice's and Bob's sources can be trusted. It would be therefore necessary to further investigate
how this essential requirement could be guaranteed in practice. Also, it would be important to take both source flaws and detector flaws into account by combining
mdiQKD with the recent security analysis reported in~\cite{kiyo_loss}. Such result, and its experimental demonstration,
would bring QKD a big step closer to achieving unconditional security.
In addition,
it would be beneficial to derive tighter finite-key security bounds
such that the post-processing data block sizes needed to achieve good performance could be reduced.

Moreover, the idea of mdiQKD can be applied as well to solve other quantum information tasks, such as the evaluation of
entanglement witness~\cite{branciard2013measurement,xu2014implementation}, randomness certification~\cite{banik2014measurement},
or to develop diQKD systems that are robust against channel losses~\cite{lim2013device}.

In summary, mdiQKD enables new scientific developments in the field of quantum optics, as well as advanced novel applications for quantum information
and quantum communication.
In a popular book ``The Code Book''~\cite{simon_book} by Simon Singh, the author
suggested that QKD will be the end point of evolution of
cryptography. In this paper, we have shown that while quantum hacking has
threatened the security of QKD, mdiQKD has now appeared
to be an important counter-measure against it, thus constituting
a major step towards
the Holy Grail of cryptography---unconditional security in
communication.

\section*{Acknowledgments}
The authors thank L. Qian, K. Tamaki for helpful discussions. We thank Z. Tang, Y. Li and Q. Zhang for allowing us to use the figures of polarisation-encoding and time-bin mdiQKD setup. Support from NSERC, the CRC program, Connaught Innovation fund and Industry Canada, the European Regional Development Fund (ERDF), the Spanish National Research and Development Program under project Terasense CSD2008- 00068
(Consolider-Ingenio 2010), and the Galician Regional Government (projects CN2012/279, CN 2012/260, consolidation of Research Units: AtlantTIC, and the program "ayudas para proyectos de investigaci\'{o}n desarrollados por investigadores emergentes"), and the Laboratory Directed Research and Development Program of Oak Ridge National Laboratory, managed by UT-Battelle, LLC, for the U. S. Department of Energy is gratefully acknowledged. F. Xu thanks the Shahid U.H. Qureshi Memorial Scholarship and the OGS VISA award for financial support.

\bibliographystyle{IEEEtran}
\bibliography{MDIreview_final}

\begin{thebibliography}{10}
\providecommand{\url}[1]{#1}
\csname url@samestyle\endcsname
\providecommand{\newblock}{\relax}
\providecommand{\bibinfo}[2]{#2}
\providecommand{\BIBentrySTDinterwordspacing}{\spaceskip=0pt\relax}
\providecommand{\BIBentryALTinterwordstretchfactor}{4}
\providecommand{\BIBentryALTinterwordspacing}{\spaceskip=\fontdimen2\font plus
\BIBentryALTinterwordstretchfactor\fontdimen3\font minus
  \fontdimen4\font\relax}
\providecommand{\BIBforeignlanguage}[2]{{%
\expandafter\ifx\csname l@#1\endcsname\relax
\typeout{** WARNING: IEEEtran.bst: No hyphenation pattern has been}%
\typeout{** loaded for the language `#1'. Using the pattern for}%
\typeout{** the default language instead.}%
\else
\language=\csname l@#1\endcsname
\fi
#2}}
\providecommand{\BIBdecl}{\relax}
\BIBdecl

\bibitem{Vernam:OTP:1926}
G.~Vernam, ``Cipher printing telegraph systems for secret wire and radio
  telegraphic communications,'' \emph{Transactions of the American Institute of
  Electrical Engineers}, vol.~45, pp. 109--115, 1926.

\bibitem{bennett1984quantum}
C.~H. Bennett and G.~Brassard, ``Quantum cryptography: Public key distribution
  and coin tossing,'' in \emph{Proceedings of IEEE International Conference on
  Computers, Systems and Signal Processing}.\hskip 1em plus 0.5em minus
  0.4em\relax Bangalore, India, 1984, pp. 175--179.

\bibitem{RSA:1978}
R.~Rivest, A.~Shamir, and L.~Adleman, ``A method for obtaining digital
  signatures and public-key cryptosystems,'' \emph{Communications of the ACM},
  vol.~21, no.~2, pp. 120--126, 1978.

\bibitem{Shor:1997}
P.~W. Shor, ``Polynomial-time algorithms for prime factorization and discrete
  logarithms on a quantum computer,'' \emph{SIAM J. Sci. Statist. Comput.},
  vol.~26, p. 1484, 1997.

\bibitem{Mayers:2001}
D.~Mayers, ``Unconditional security in quantum cryptography,'' \emph{Journal of
  the ACM}, vol.~48, no.~3, pp. 351--406, 2001.

\bibitem{Lo:1999}
H.-K. Lo and H.~F. Chau, ``Unconditional security of quantum key distribution
  over arbitrarily long distances,'' \emph{Science}, vol. 283, no. 5410, pp.
  2050--2056, 1999.

\bibitem{Shor:2000}
P.~W. Shor and J.~Preskill, ``Simple proof of security of the bb84 quantum key
  distribution protocol,'' \emph{Phys. Rev. Lett.}, vol.~85, no.~2, pp.
  441--444, 2000.

\bibitem{GLLP:2004}
D.~Gottesman, H.-K. Lo, N.~L\"utkenhaus, and J.~Preskill, ``Security of quantum
  key distribution with imperfect devices,'' \emph{Quant. Inf. Comput.},
  vol.~4, pp. 325--360, 2004.

\bibitem{biham05}
E.~Biham, M.~Boyer, P.~O. Boykin, T.~Mor, and V.~Roychowdhury, ``A proof of the
  security of quantum key distribution,'' \emph{Journal of Cryptology},
  vol.~19, pp. 381--439, 2006.

\bibitem{Yi:timeshift:2008}
Y.~Zhao, C.~Fung, B.~Qi, C.~Chen, and H.-K. Lo, ``Quantum hacking: Experimental
  demonstration of time-shift attack against practical quantum-key-distribution
  systems,'' \emph{Physical Review A}, vol.~78, p. 042333, 2008.

\bibitem{Qi:timeshift:2007}
B.~Qi, C.~Fung, H.-K. Lo, and X.~Ma, ``Time-shift attack in practical quantum
  cryptosystems,'' \emph{Quant. Inf. Comput.}, vol.~7, p.~73, 2007.

\bibitem{Hadfield:2009}
R.~Hadfield, ``Single-photon detectors for optical quantum information
  applications,'' \emph{Nature Photonicss}, vol.~3, no.~12, pp. 696--705, 2009.

\bibitem{Makarov:fakestate:2006}
V.~Makarov, A.~Anisimov, and J.~Skaar, ``Effects of detector efficiency
  mismatch on security of quantum cryptosystems,'' \emph{Physical Review A},
  vol.~74, no. 022313, 2006.

\bibitem{Lars:nature:2010}
L.~Lydersen, C.~Wiechers, C.~Wittmann, D.~Elser, J.~Skaar, and V.~Makarov,
  ``Hacking commercial quantum cryptography systems by tailored bright
  illumination,'' \emph{Nature Photonics}, vol.~4, no.~10, pp. 686--689, 2010.

\bibitem{yuan2010avoiding}
Z.~Yuan, J.~Dynes, and A.~Shields, ``Avoiding the blinding attack in qkd,''
  \emph{Nature Photonicss}, vol.~4, no.~12, pp. 800--801, 2010.

\bibitem{Lars:avoiding:2010}
L.~Lydersen, C.~Wiechers, C.~Wittmann, D.~Elser, J.~Skaar, and V.~Makarov,
  ``Reply to ``avoiding the blinding attack in qkd'','' \emph{Nature
  Photonics}, vol.~4, no.~12, p. 801, 2010.

\bibitem{Gerhardt:2010}
I.~Gerhardt, Q.~Liu, A.~Lamas-Linares, J.~Skaar, C.~Kurtsiefer, and V.~Makarov,
  ``Perfect eavesdropping on a quantum cryptography system,'' \emph{Nature
  Communications}, vol.~2, p. 349, 2011.

\bibitem{timeinfo}
A.~Lamas-Linares and C.~Kurtsiefer, ``Breaking a quantum key distribution
  system through a timing side channel,'' \emph{Opt. Express}, vol.~15, pp.
  9388--9393, 2007.

\bibitem{Xu:phaseremapping:2010}
F.~Xu, B.~Qi, and H.-K. Lo, ``Experimental demonstration of phase-remapping
  attack in a practical quantum key distribution system,'' \emph{New Journal of
  Physics}, vol.~12, p. 113026, 2010.

\bibitem{weier2011quantum}
H.~Weier, H.~Krauss, M.~Rau, M.~Fuerst, S.~Nauerth, and H.~Weinfurter,
  ``Quantum eavesdropping without interception: An attack exploiting the dead
  time of single photon detectors,'' \emph{New Journal of Physics}, vol.~13, p.
  073024, 2011.

\bibitem{jain2011device}
N.~Jain, C.~Wittmann, L.~Lydersen, C.~Wiechers, D.~Elser, C.~Marquardt,
  V.~Makarov, and G.~Leuchs, ``Device calibration impacts security of quantum
  key distribution,'' \emph{Physical Review Letters}, vol. 107, no.~11, p.
  110501, 2011.

\bibitem{li2011attacking}
H.-W. Li, S.~Wang, J.-Z. Huang, W.~Chen, Z.-Q. Yin, F.-Y. Li, Z.~Zhou, D.~Liu,
  Y.~Zhang, G.-C. Guo \emph{et~al.}, ``Attacking a practical
  quantum-key-distribution system with wavelength-dependent beam-splitter and
  multiwavelength sources,'' \emph{Physical Review A}, vol.~84, no.~6, p.
  062308, 2011.

\bibitem{tang2013source}
Y.-L. Tang, H.-L. Yin, X.~Ma, C.-H.~F. Fung, Y.~Liu, H.-L. Yong, T.-Y. Chen,
  C.-Z. Peng, Z.-B. Chen, and J.-W. Pan, ``Source attack of decoy-state quantum
  key distribution using phase information,'' \emph{Physical Review A},
  vol.~88, no.~2, p. 022308, 2013.

\bibitem{jouguet2013preventing}
P.~Jouguet, S.~Kunz-Jacques, and E.~Diamanti, ``Preventing calibration attacks
  on the local oscillator in continuous-variable quantum key distribution,''
  \emph{Physical Review A}, vol.~87, no.~6, p. 062313, 2013.

\bibitem{jiang2013intrinsic}
M.-S. Jiang, S.-H. Sun, G.-Z. Tang, X.-C. Ma, C.-Y. Li, and L.-M. Liang,
  ``Intrinsic imperfection of self-differencing single-photon detectors harms
  the security of high-speed quantum cryptography systems,'' \emph{Physical
  Review A}, vol.~88, no.~6, p. 062335, 2013.

\bibitem{PhysRevLett.112.070503}
A.~N. Bugge, S.~Sauge, A.~M.~M. Ghazali, J.~Skaar, L.~Lydersen, and V.~Makarov,
  ``Laser damage helps the eavesdropper in quantum cryptography,'' \emph{Phys.
  Rev. Lett.}, vol. 112, p. 070503, 2014.

\bibitem{kiyo_loss}
\BIBentryALTinterwordspacing
K.~Tamaki, M.~Curty, G.~Kato, H.-K. Lo, and K.~Azuma, ``Loss-tolerant quantum
  cryptography with imperfect sources,'' \emph{Phys. Rev. A}, vol.~90, p.
  052314, Nov 2014. [Online]. Available:
  \url{http://link.aps.org/doi/10.1103/PhysRevA.90.052314}
\BIBentrySTDinterwordspacing

\bibitem{xu2014experimental}
F.~Xu, S.~Sajeed, S.~Kaiser, Z.~Tang, L.~Qian, V.~Makarov, and H.-K. Lo,
  ``Experimental quantum key distribution with source flaws and tight
  finite-key analysis,'' \emph{arXiv preprint arXiv:1408.3667}, 2014.

\bibitem{Fred:Proof:2009}
C.~Fung, K.~Tamaki, B.~Qi, H.-K. Lo, and X.~Ma, ``Security proof of quantum key
  distribution with detection efficiency mismatch,'' \emph{Quant. Inf.
  Comput.}, vol.~9, p. 131, 2009.

\bibitem{PhysRevA.82.032337}
O.~Mar\o{}y, L.~Lydersen, and J.~Skaar, ``Security of quantum key distribution
  with arbitrary individual imperfections,'' \emph{Phys. Rev. A}, vol.~82, p.
  032337, 2010.

\bibitem{ferreira2012real}
T.~Ferreira~da Silva, G.~B. Xavier, G.~P. Tempor{\~a}o, and J.~P. von~der Weid,
  ``Real-time monitoring of single-photon detectors against eavesdropping in
  quantum key distribution systems,'' \emph{Optics express}, vol.~20, no.~17,
  pp. 18\,911--18\,924, 2012.

\bibitem{yuan2011resilience}
Z.~Yuan, J.~Dynes, and A.~Shields, ``Resilience of gated avalanche photodiodes
  against bright illumination attacks in quantum cryptography,'' \emph{Applied
  physics letters}, vol.~98, no.~23, pp. 231\,104--231\,104, 2011.

\bibitem{mayers_diQKD}
D.~Mayers and A.~C.-C. Yao, ``Quantum cryptography with imperfect apparatus,''
  in \emph{Proceedings of the 39th Annual Symposium on Foundations of Computer
  Science}.\hskip 1em plus 0.5em minus 0.4em\relax IEEE Computer Society Press,
  1998, pp. 503--509.

\bibitem{acin2007device}
A.~Ac{\'\i}n, N.~Brunner, N.~Gisin, S.~Massar, S.~Pironio, and V.~Scarani,
  ``Device-independent security of quantum cryptography against collective
  attacks,'' \emph{Physical Review Letters}, vol.~98, no.~23, p. 230501, 2007.

\bibitem{diQKD3}
B.~W. Reichardt, F.~Unger, and U.~Vazirani, ``Classical command of quantum
  systems,'' \emph{Nature}, vol. 496, pp. 456--460, 2013.

\bibitem{qbitamp1}
N.~Gisin, S.~Pironio, and N.~Sangouard, ``Proposal for implementing
  device-independent quantum key distribution based on a heralded qubit
  amplifier,'' \emph{Phys. Rev. Lett.}, vol. 105, p. 070501, 2010.

\bibitem{qbitamp2}
M.~Curty and T.~Moroder, ``Heralded-qubit amplifiers for practical
  device-independent quantum key distribution,'' \emph{Phys. Rev. A}, vol.~84,
  p. 010304(R), 2011.

\bibitem{Lo:MDIQKD}
H.-K. Lo, M.~Curty, and B.~Qi, ``Measurement-device-independent quantum key
  distribution,'' \emph{Physical Review Letters}, vol. 108, no.~13, p. 130503,
  2012.

\bibitem{biham1996quantum}
E.~Biham, B.~Huttner, and T.~Mor, ``Quantum cryptographic network based on
  quantum memories,'' \emph{Physical Review A}, vol.~54, no.~4, p. 2651, 1996.

\bibitem{zukowski1993event}
M.~Zukowski, A.~Zeilinger, M.~Horne, and A.~K. Ekert,
  ``{``Event-ready-detectors"} {Bell} experiment via entanglement swapping,''
  \emph{Physical Review Letters}, vol.~71, no.~26, pp. 4287--4290, 1993.

\bibitem{pan1998experimental}
J.-W. Pan, D.~Bouwmeester, H.~Weinfurter, and A.~Zeilinger, ``Experimental
  entanglement swapping: Entangling photons that never interacted,''
  \emph{Physical Review Letters}, vol.~80, no.~18, p. 3891, 1998.

\bibitem{inamori2002security}
H.~Inamori, ``Security of practical time-reversed {EPR} quantum key
  distribution,'' \emph{Algorithmica}, vol.~34, pp. 340--365, 2002.

\bibitem{braunstein2012side}
S.~L. Braunstein and S.~Pirandola, ``Side-channel-free quantum key
  distribution,'' \emph{Physical Review Letters}, vol. 108, no.~13, p. 130502,
  2012.

\bibitem{Hwang:2003}
W.~Hwang, ``Quantum key distribution with high loss: Toward global secure
  communication,'' \emph{Physical Review Letters}, vol.~91, p. 57901, 2003.

\bibitem{Lo:2005}
H.-K. Lo, X.~Ma, and K.~Chen, ``Decoy state quantum key distribution,''
  \emph{Physical Review Letters}, vol.~94, p. 230504, 2005.

\bibitem{Wang:2005}
X.~Wang, ``Beating the photon-number-splitting attack in practical quantum
  cryptography,'' \emph{Physical Review Letters}, vol.~94, p. 230503, 2005.

\bibitem{branciard2012one}
C.~Branciard, E.~G. Cavalcanti, S.~P. Walborn, V.~Scarani, and H.~M. Wiseman,
  ``One-sided device-independent quantum key distribution: Security,
  feasibility, and the connection with steering,'' \emph{Physical Review A},
  vol.~85, no.~1, p. 010301, 2012.

\bibitem{yin2013measurement}
Z.-Q. Yin, C.-H.~F. Fung, X.~Ma, C.-M. Zhang, H.-W. Li, W.~Chen, S.~Wang, G.-C.
  Guo, and Z.-F. Han, ``Measurement-device-independent quantum key distribution
  with uncharacterized qubit sources,'' \emph{Physical Review A}, vol.~88,
  no.~6, p. 062322, 2013.

\bibitem{yin2014mismatched}
Z.-Q. Yin, C.-H.~F. Fung, X.~Ma, C.-M. Zhang \emph{et~al.}, ``Mismatched-basis
  statistics enable quantum key distribution with uncharacterized qubit
  sources,'' \emph{arXiv preprint arXiv:1407.1924}, 2014.

\bibitem{lo2005efficient}
H.-K. Lo, H.-F. Chau, and M.~Ardehali, ``Efficient quantum key distribution
  scheme and a proof of its unconditional security,'' \emph{Journal of
  Cryptology}, vol.~18, no.~2, pp. 133--165, 2005.

\bibitem{Tittel:2012:MDI:exp}
A.~Rubenok, J.~A. Slater, P.~Chan, I.~Lucio-Martinez, and W.~Tittel,
  ``Real-world two-photon interference and proof-of-principle quantum key
  distribution immune to detector attacks,'' \emph{Phys. Rev. Lett.}, vol. 111,
  p. 130501, 2013.

\bibitem{da2012proof}
T.~Ferreira~da Silva, D.~Vitoreti, G.~B. Xavier, G.~C. do~Amaral, G.~P.
  Tempor\~ao, and J.~P. von~der Weid, ``Proof-of-principle demonstration of
  measurement-device-independent quantum key distribution using polarization
  qubits,'' \emph{Phys. Rev. A}, vol.~88, p. 052303, 2013.

\bibitem{Liu:2012:MDI:exp}
Y.~Liu, T.-Y. Chen, L.-J. Wang, H.~Liang, G.-L. Shentu, J.~Wang, K.~Cui, H.-L.
  Yin, N.-L. Liu, L.~Li \emph{et~al.}, ``Experimental
  measurement-device-independent quantum key distribution,'' \emph{Phys. Rev.
  Lett.}, vol. 111, p. 130502, 2013.

\bibitem{zhiyuan:experiment:2013}
Z.~Tang, Z.~Liao, F.~Xu, B.~Qi, L.~Qian, and H.-K. Lo, ``Experimental
  demonstration of polarization encoding measurement-device-independent quantum
  key distribution,'' \emph{Physical Review Letters}, vol. 112, no.~19, p.
  190503, 2014.

\bibitem{dixon2008gigahertz}
A.~Dixon, Z.~Yuan, J.~Dynes, A.~Sharpe, and A.~Shields, ``Gigahertz decoy
  quantum key distribution with 1 mbit/s secure key rate,'' \emph{Optics
  Express}, vol.~16, no.~23, pp. 18\,790--18\,979, 2008.

\bibitem{rosenberg2009practical}
D.~Rosenberg, C.~Peterson, J.~Harrington, P.~Rice, N.~Dallmann, K.~Tyagi,
  K.~McCabe, S.~Nam, B.~Baek, R.~Hadfield \emph{et~al.}, ``Practical
  long-distance quantum key distribution system using decoy levels,'' \emph{New
  Journal of Physics}, vol.~11, no.~4, p. 045009, 2009.

\bibitem{curty2014finite}
M.~Curty, F.~Xu, W.~Cui, C.~C.~W. Lim, K.~Tamaki, and H.-K. Lo, ``Finite-key
  analysis for measurement-device-independent quantum key distribution,''
  \emph{Nature communications}, vol.~5, p. 3732, 2014.

\bibitem{ma2012statistical}
X.~Ma, C.-H.~F. Fung, and M.~Razavi, ``Statistical fluctuation analysis for
  measurement-device-independent quantum key distribution,'' \emph{Physical
  Review A}, vol.~86, no.~5, p. 052305, 2012.

\bibitem{wang2013three}
X.-B. Wang, ``Three-intensity decoy-state method for device-independent quantum
  key distribution with basis-dependent errors,'' \emph{Physical Review A},
  vol.~87, no.~1, p. 012320, 2013.

\bibitem{sun2013practical}
S.-H. Sun, M.~Gao, C.-Y. Li, and L.-M. Liang, ``Practical decoy-state
  measurement-device-independent quantum key distribution,'' \emph{Physical
  Review A}, vol.~87, no.~5, p. 052329, 2013.

\bibitem{Feihu:practical}
F.~Xu, M.~Curty, B.~Qi, and H.-K. Lo, ``Practical aspects of
  measurement-device-independent quantum key distribution,'' \emph{New Journal
  of Physics}, vol.~15, no.~11, p. 113007, 2013.

\bibitem{PhysRevA.89.052333}
F.~Xu, H.~Xu, and H.-K. Lo, ``Protocol choice and parameter optimization in
  decoy-state measurement-device-independent quantum key distribution,''
  \emph{Phys. Rev. A}, vol.~89, p. 052333, 2014.

\bibitem{renner2005security}
R.~Renner, ``Security of quantum key distribution,'' \emph{PhD Thesis, ETH
  No.16242, arXiv: quant-ph/0512258}, 2005.

\bibitem{scarani2008quantum}
V.~Scarani and R.~Renner, ``Quantum cryptography with finite resources:
  Unconditional security bound for discrete-variable protocols with one-way
  postprocessing,'' \emph{Physical Review Letters}, vol. 100, no.~20, p.
  200501, 2008.

\bibitem{QKD:Scarani:2009}
V.~Scarani, H.~Bechmann-Pasquinucci, N.~Cerf, M.~Du{\v{s}}ek,
  N.~L{\"u}tkenhaus, and M.~Peev, ``The security of practical quantum key
  distribution,'' \emph{Reviews of Modern Physics}, vol.~81, no.~3, p. 1301,
  2009.

\bibitem{tomamichel2012tight}
M.~Tomamichel, C.~C.~W. Lim, N.~Gisin, and R.~Renner, ``Tight finite-key
  analysis for quantum cryptography,'' \emph{Nature Communications}, vol.~3, p.
  634, 2012.

\bibitem{hayashi2014security}
M.~Hayashi and R.~Nakayama, ``Security analysis of the decoy method with the
  bennett--brassard 1984 protocol for finite key lengths,'' \emph{New Journal
  of Physics}, vol.~16, no.~6, p. 063009, 2014.

\bibitem{lim2014concise}
C.~C.~W. Lim, M.~Curty, N.~Walenta, F.~Xu, and H.~Zbinden, ``Concise security
  bounds for practical decoy-state quantum key distribution,'' \emph{Physical
  Review A}, vol.~89, no.~2, p. 022307, 2014.

\bibitem{SongPhysRevA.86.022332}
T.-T. Song, Q.-Y. Wen, F.-Z. Guo, and X.-Q. Tan, ``Finite-key analysis for
  measurement-device-independent quantum key distribution,'' \emph{Phys. Rev.
  A}, vol.~86, p. 022332, 2012.

\bibitem{ben2005universal}
M.~Ben-Or, M.~Horodecki, D.~W. Leung, D.~Mayers, and J.~Oppenheim, ``The
  universal composable security of quantum key distribution,'' in \emph{Theory
  of Cryptography: Second Theory of Cryptography Conference, TCC 2005}.\hskip
  1em plus 0.5em minus 0.4em\relax Springer, 2005, pp. 386--406.

\bibitem{renner2005universally}
R.~Renner and R.~K{\"o}nig, ``Universally composable privacy amplification
  against quantum adversaries,'' in \emph{Theory of Cryptography: Second Theory
  of Cryptography Conference, TCC 2005}.\hskip 1em plus 0.5em minus 0.4em\relax
  Springer, 2005, pp. 407--425.

\bibitem{tamaki2012phase}
K.~Tamaki, H.-K. Lo, C.-H.~F. Fung, and B.~Qi, ``Phase encoding schemes for
  measurement-device-independent quantum key distribution with basis-dependent
  flaw,'' \emph{Physical Review A}, vol.~85, no.~4, p. 042307, 2012.

\bibitem{ma2012alternative}
X.~Ma and M.~Razavi, ``Alternative schemes for measurement-device-independent
  quantum key distribution,'' \emph{Physical Review A}, vol.~86, no.~6, p.
  062319, 2012.

\bibitem{chan2014modeling}
P.~Chan, J.~Slater, I.~Lucio-Martinez, A.~Rubenok, and W.~Tittel, ``Modeling a
  measurement-device-independent quantum key distribution system,''
  \emph{Optics Express}, vol.~22, no.~11, pp. 12\,716--12\,736, 2014.

\bibitem{stucki2009high}
D.~Stucki, N.~Walenta, F.~Vannel, R.~Thew, N.~Gisin, H.~Zbinden, S.~Gray,
  C.~Towery, and S.~Ten, ``High rate, long-distance quantum key distribution
  over 250 km of ultra low loss fibres,'' \emph{New Journal of Physics},
  vol.~11, p. 075003, 2009.

\bibitem{marsili2013detecting}
F.~Marsili, V.~Verma, J.~Stern, S.~Harrington, A.~Lita, T.~Gerrits,
  I.~Vayshenker, B.~Baek, M.~Shaw, R.~Mirin \emph{et~al.}, ``Detecting single
  infrared photons with 93\% system efficiency,'' \emph{Nature Photonics},
  vol.~7, no.~3, pp. 210--214, 2013.

\bibitem{abruzzo2014measurement}
S.~Abruzzo, H.~Kampermann, and D.~Bru{\ss}, ``Measurement-device-independent
  quantum key distribution with quantum memories,'' \emph{Physical Review A},
  vol.~89, no.~1, p. 012301, 2014.

\bibitem{panayi2014memory}
C.~Panayi, M.~Razavi, X.~Ma, and N.~L{\"u}tkenhaus, ``Memory-assisted
  measurement-device-independent quantum key distribution,'' \emph{New Journal
  of Physics}, vol.~16, no.~4, p. 043005, 2014.

\bibitem{xu2013long}
F.~Xu, B.~Qi, Z.~Liao, and H.-K. Lo, ``Long distance
  measurement-device-independent quantum key distribution with entangled photon
  sources,'' \emph{Applied Physics Letters}, vol. 103, no.~6, p. 061101, 2013.

\bibitem{li2014continuous}
Z.~Li, Y.-C. Zhang, F.~Xu, X.~Peng, and H.~Guo, ``Continuous-variable
  measurement-device-independent quantum key distribution,'' \emph{Physical
  Review A}, vol.~89, no.~5, p. 052301, 2014.

\bibitem{pirandola2013continuous}
S.~Pirandola, C.~Ottaviani, G.~Spedalieri, C.~Weedbrook, and S.~L. Braunstein,
  ``Continuous-variable quantum cryptography with untrusted relays,''
  \emph{arXiv preprint arXiv:1312.4104}, 2013.

\bibitem{PhysRevA.89.042335}
X.-C. Ma, S.-H. Sun, M.-S. Jiang, M.~Gui, and L.-M. Liang, ``Gaussian-modulated
  coherent-state measurement-device-independent quantum key distribution,''
  \emph{Phys. Rev. A}, vol.~89, p. 042335, 2014.

\bibitem{mizutani2014measurement}
A.~Mizutani, K.~Tamaki, R.~Ikuta, T.~Yamamoto, and N.~Imoto,
  ``Measurement-device-independent quantum key distribution for
  scarani-acin-ribordy-gisin 04 protocol,'' \emph{Scientific reports}, vol.~4,
  p. 5236, 2014.

\bibitem{PhysRevA.88.052332}
Q.~Wang and X.-B. Wang, ``Efficient implementation of the decoy-state
  measurement-device-independent quantum key distribution with heralded
  single-photon sources,'' \emph{Phys. Rev. A}, vol.~88, p. 052332, 2013.

\bibitem{li2014measurement}
M.~Li, C.-M. Zhang, Z.-Q. Yin, W.~Chen, S.~Wang, G.-C. Guo, and Z.-F. Han,
  ``Measurement-device-independent quantum key distribution with modified
  coherent state,'' \emph{Optics letters}, vol.~39, no.~4, pp. 880--883, 2014.

\bibitem{new_mdiQKDexp}
Y.-L. Tang, H.-L. Yin, S.-J. Chen, Y.~Liu, W.-J. Zhang, X.~Jiang, L.~Zhang,
  J.~Wang, L.-X. You, J.-Y. Guan \emph{et~al.},
  ``Measurement-device-independent quantum key distribution over 200 km,''
  \emph{Physical Review Letters}, vol. 113, no.~19, p. 190501, 2014.

\bibitem{6920009}
Y.-L. Tang, H.-L. Yin, S.-J. Chen, Y.~Liu, W.-J. Zhang, X.~Jiang, L.~Zhang,
  J.~Wang, L.-X. You, J.-Y. Guan, D.-X. Yang, Z.~Wang, H.~Liang, Z.~Zhang,
  N.~Zhou, X.~Ma, T.~Chen, Q.~Zhang, and J.-W. Pan, ``Field test of
  measurement-device-independent quantum key distribution,'' \emph{Selected
  Topics in Quantum Electronics, IEEE Journal of}, vol.~PP, no.~99, pp. 1--1,
  2014.

\bibitem{frohlich2013quantum}
B.~Fr{\"o}hlich, J.~F. Dynes, M.~Lucamarini, A.~W. Sharpe, Z.~Yuan, and A.~J.
  Shields, ``A quantum access network,'' \emph{Nature}, vol. 501, no. 7465, pp.
  69--72, 2013.

\bibitem{hughes2013network}
R.~J. Hughes, J.~E. Nordholt, K.~P. McCabe, R.~T. Newell, C.~G. Peterson, and
  R.~D. Somma, ``Network-centric quantum communications with application to
  critical infrastructure protection,'' \emph{arXiv preprint arXiv:1305.0305},
  2013.

\bibitem{branciard2013measurement}
C.~Branciard, D.~Rosset, Y.-C. Liang, and N.~Gisin,
  ``Measurement-device-independent entanglement witnesses for all entangled
  quantum states,'' \emph{Physical Review Letters}, vol. 110, no.~6, p. 060405,
  2013.

\bibitem{xu2014implementation}
P.~Xu, X.~Yuan, L.-K. Chen, H.~Lu, X.-C. Yao, X.~Ma, Y.-A. Chen, and J.-W. Pan,
  ``Implementation of a measurement-device-independent entanglement witness,''
  \emph{Physical Review Letters}, vol. 112, no.~14, p. 140506, 2014.

\bibitem{banik2014measurement}
M.~Banik, ``Measurement-device-independent randomness certification by local
  entangled states,'' \emph{arXiv preprint arXiv:1401.1338}, 2014.

\bibitem{lim2013device}
C.~C.~W. Lim, C.~Portmann, M.~Tomamichel, R.~Renner, and N.~Gisin,
  ``Device-independent quantum key distribution with local bell test,''
  \emph{Physical Review X}, vol.~3, no.~3, p. 031006, 2013.

\bibitem{simon_book}
S.~Singh, \emph{The Code Book: The Science of Secrecy from Ancient Egypt to
  Quantum Cryptography}.\hskip 1em plus 0.5em minus 0.4em\relax Doubleday, New
  York, 1999.

\end{thebibliography}

\begin{IEEEbiography}[{\includegraphics[width=1in,height=1.35in,clip,keepaspectratio]{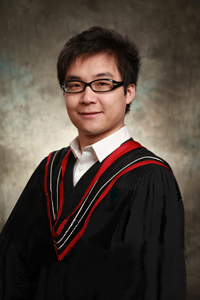}}]{Feihu Xu}
received the B.Sc. degree (with Hons.)
from the University of Science and Technology of
China, Hefei, China, in 2009. After that, he went
to the Department of Electrical and Computer Engineering
(ECE), University of Toronto, Toronto, ON,
Canada, for graduate study, where he received the
M.A.Sc. and Ph.D. degrees in 2011 and 2014, respectively.
He moved to the Research Laboratory of
Electronics, Massachusetts Institute of Technology
(MIT) as a Postdoctoral Associate in January 2015.
His current research interest lies in the broad area of
quantum information processing, particularly in quantum communication, quantum
cryptography, quantum random number generation, multi-photon quantum
interference and boson sampling.
\end{IEEEbiography}

\begin{IEEEbiography}[{\includegraphics[width=1in,height=1.35in,clip,keepaspectratio]{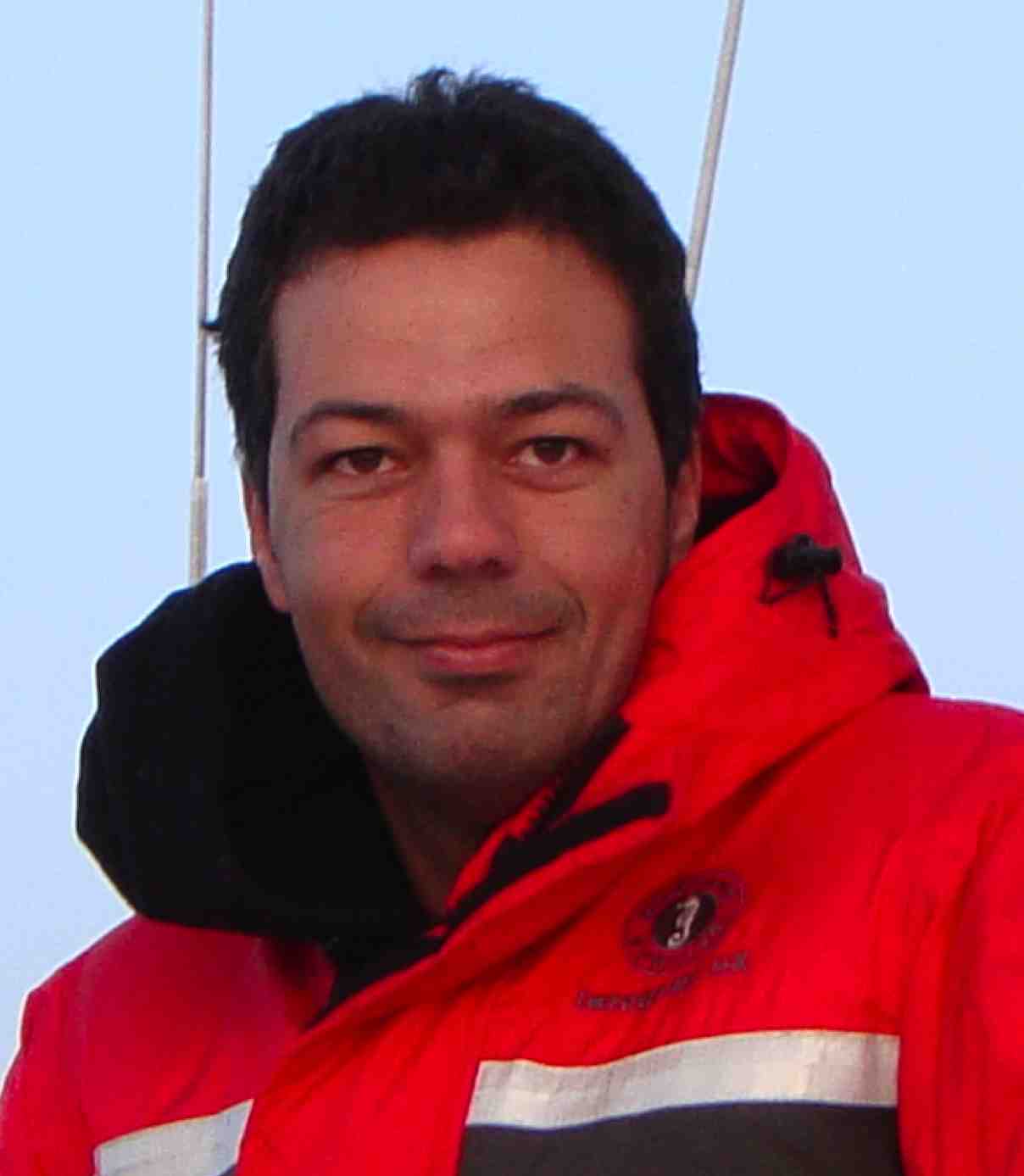}}]{Marcos Curty}
received his M.Sc. and Ph.D in Telecommunication Engineering from Vigo University in 1999 and 2004 respectively. In 2001 he joined the
Quantum Information Theory Group at the University of Erlangen-N\"urnberg and he obtained a Ph.D in Physics (2006) under the supervision of Prof. Norbert L\"utkenhaus. After postdoc positions at Toronto University and at the Institute for Quantum Computing, Waterloo University, he joined the Department of Electronic and Communications Engineering at Zaragoza University as Assistant Professor. Currently he is an Associate Professor in the Theory of Signal and Communications Department at Vigo University. His research interest lies in quantum information processing, particularly quantum cryptography.
\end{IEEEbiography}

\begin{IEEEbiography}[{\includegraphics[width=1in,height=1.35in,clip,keepaspectratio]{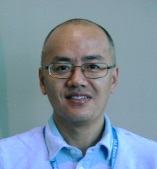}}]{Bing Qi}
is a Research Scientist at Oak Ridge National Laboratory (ORNL). He received a Ph.D. in Optics in 1996 from Dalian University of Technology and joined the ORNL staff in 2013. Before joining ORNL, he was a Senior Research Associate in the Quantum Communication group and in the Emerging Fiber-Optic Technologies group, Dept. of Electrical and Computer Engineering, University of Toronto. Dr. Qi's research interests include quantum communication and optical sensing.
\end{IEEEbiography}

\begin{IEEEbiography}[{\includegraphics[width=1in,height=1.35in,clip,keepaspectratio]{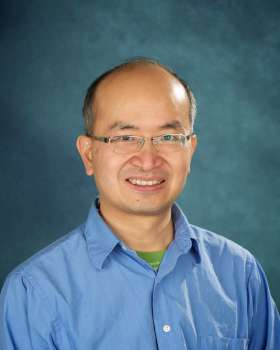}}]{Hoi-Kwong Lo} is a Full Professor at the Center for Quantum Information and Quantum Control (CQIQC), the Department of Physics and the Department of Electrical and Computer Engineering, at the University of Toronto. His research
interest lies in quantum information processing, particularly quantum cryptography. He received his B.A. in Mathematics from Trinity College, Cambridge University in 1989 and his M.Sc. and Ph.D. in Physics from Caltech in 1991 and 1994 respectively. Prof. Lo had six years of industrial research experience at Hewlett-Packard Lab., Bristol UK and also as the Chief Scientist and Senior Vice President, R\&D, of MagiQ Technologies, Inc., New York. He was a co-founder of a leading journal Quantum Information and Computation (QIC).
\end{IEEEbiography}

\end{document}